\documentstyle[12pt]{article}
\parskip=\medskipamount                 
\thicklines                     
\newcommand{\bim}[6]{\bibitem{#1}#2, {\em #3\/}$\;${\bf
#4}$\;$(#5)$\;${#6}.}
\def\IR{\relax{\rm I\kern-.18em R}}
\def\ZZ{\relax{\sf Z\kern-.4em Z}}
\def\a{\alpha} \def\b{\beta}   \def\e{\epsilon} 
\def\G{\Gamma}  
   
   \def\cD{{\cal D}}
   
 \def\cK{{\cal K}} \def\cL{{\cal L}} \def\cM{{\cal M}}
 \def\cO{{\cal O}}  

\def\etc{{\it etc\/}}

\newtheorem{proposition}{Proposition}[section]

\newtheorem{conjecture}{Conjecture}[section]
\newtheorem{lemma}{Lemma}[section]

\catcode`\@=11

\newif\if@fewtab\@fewtabtrue

\catcode`\@=11

\newif\if@fewtab\@fewtabtrue

{\count255=\time\divide\count255 by 60
\xdef\hourmin{\number\count255}
\multiply\count255 by-60\advance\count255 by\time
\xdef\hourmin{\hourmin:\ifnum\count255<10 0\fi\the\count255}}
\def\ps@draft{\let\@mkboth\@gobbletwo
    \def\@oddhead{}
    \def\@oddfoot
       {\hbox to 7 cm{$\scriptstyle Draft\ version:\ \draftdate$
       \hfil}\hskip -7cm\hfil\rm\thepage \hfil}
    \def\@evenhead{}\let\@evenfoot\@oddfoot}

\def\ceqno{\global\@fewtabfalse
    \ifcase\@eqcnt \def\@tempa{& & &}\or \def\@tempa{& &}
      \or \def\@tempa{&}
      \or\def\@tempa{}\fi\@tempa
{\rm(\theequation)}}

\def\aeqno#1{\global\@fewtabfalse
    \ifcase\@eqcnt \def\@tempa{& & &}\or \def\@tempa{& &}
      \or \def\@tempa{&}
      \or\def\@tempa{}\fi\@tempa
{\rm(\theequation,#1)}}

\def\label#1{\ifnum\draftcontrol=1
 \global\def\draftnote{$\scriptstyle #1$}\fi
 \@bsphack\if@filesw {\let\thepage\relax
   \def\protect{\noexpand\noexpand\noexpand}%
\xdef\@gtempa{\write\@auxout{\string
      \newlabel{#1}{{\@currentlabel}{\thepage}}}}}\@gtempa
   \if@nobreak \ifvmode\nobreak\fi\fi\fi
  \@esphack}

\def\alabel#1#2{\label{#1}\global\@fewtabfalse
    \ifcase\@eqcnt \def\@tempa{& & &}\or \def\@tempa{& &}
      \or \def\@tempa{&}
      \or\def\@tempa{}\fi\@tempa
{\hbox to 3cm{\phantom{\rm(\theequation,#2)}
\draftnote \hfil}\hskip -3cm {\rm(\theequation,#2)}}}

\def\clabel#1{\label{#1}\global\@fewtabfalse
    \ifcase\@eqcnt \def\@tempa{& & &}\or \def\@tempa{& &}
      \or \def\@tempa{&}
      \or\def\@tempa{}\fi\@tempa
{\hbox to 3cm{\phantom{\rm(\theequation)}
\draftnote \hfil}\hskip -3cm{\rm(\theequation)}}}

\def\eqnarray{\def\draftnote{{}}\global\@fewtabtrue
\stepcounter{equation}\let\@currentlabel=\theequation
\global\@eqnswtrue
\global\@eqcnt\z@\tabskip\@centering\let\\=\@eqncr
$$\halign to \displaywidth\bgroup\@eqnsel\hskip\@centering\@eqcnt\z@
  $\displaystyle\tabskip\z@{##}$&\global\@eqcnt\@ne
  \hskip 1\arraycolsep \hfil${##}$\hfil
  &\global\@eqcnt\tw@ \hskip 1\arraycolsep
$\displaystyle\tabskip\z@{##}$
\hfil  \tabskip\@centering&\global\@eqcnt\thr@@\llap{##}\tabskip\z@
\cr}

\def\endeqnarray{\@@eqncr\egroup
      \global\advance\c@equation\m@ne$$\global\@ignoretrue}

\def\@eqnnum{\hbox to 3cm{\phantom{\rm(\theequation)} \draftnote
                         \hfil}\hskip -3cm {\rm(\theequation)}}

\def\@@eqncr{\let\@tempa\relax
    \ifcase\@eqcnt \def\@tempa{& & &}\or \def\@tempa{& &}
      \or \def\@tempa{&}
      \or\def\@tempa{}
\fi\@tempa
\if@eqnsw
\if@fewtab\@eqnnum\fi
\stepcounter{equation}\fi\global
\@eqnswtrue\global\@eqcnt\z@\global\@fewtabtrue\cr}

\def\draftcite#1{\ifnum\draftcontrol=1#1\else{}\fi}

\def\@lbibitem[#1]#2{\item{}\hskip -3cm \hbox to 2cm
{\hfil$\scriptstyle\draftcite{#2}$}\hskip
1cm[\@biblabel{#1}]\if@filesw
     {\def\protect##1{\string ##1\space}\immediate
      \write\@auxout{\string\bibcite{#2}{#1}}}\fi\ignorespaces}

\def\@bibitem#1{\item\hskip -3cm \hbox to 2cm
{\hfil $\scriptstyle\draftcite{#1}$}\hskip 1cm
\if@filesw \immediate\write\@auxout
       {\string\bibcite{#1}{\the\value{\@listctr}}}\fi\ignorespaces}

\def\nsection#1{\section{#1}\setcounter{equation}{0}}
\def\nappendix#1{\def\thesection{A#1}\section*{Appendix #1}
\def\theequation{{A#1.\arabic{equation}}}
\def\theproposition{{A#1.\arabic{proposition}}}
\setcounter{equation}{0}
\setcounter{proposition}{0}}

\def\draftdate{\number\month/\number\day/\number\year\ \ \ \hourmin }

\global\def\draftcontrol{0}
\catcode`\@=12

\def\theequation{{\thesection.\arabic{equation}}}

\def\qq{\begin{eqnarray}}
\def\qqq{\end{eqnarray}}

\newlength{\shiftwidth}
\def\shift#1{&&\hbox to \shiftwidth{\hfill $\displaystyle#1$}}
\newlength{\sshiftwidth}
\def\sshift#1{\lefteqn{\hbox to
\sshiftwidth{\hfill$\displaystyle#1$}}}
\def\llefteqn#1{\hbox to 0pt{$\displaystyle #1 $\hss}\hspace*{1in}}

\def\Im{\mathop{{\rm Im}}\nolimits}
\def\Re{\mathop{{\rm Re}}\nolimits}

\def\etc{{\it etc.\ }}
\def\ie{{\it i.e.\ }}
\def\eg{{\it e.g.\ }}
\def\rhs{r.h.s.\ }

\def\Tr{\mathop{{\rm Tr}}\nolimits}
\def\Vol{\mathop{{\rm Vol}}\nolimits}

\def\zaml{Z_{\{\a\}}(M,\cL;k)}
\def\zcaml{Z_{\{\a\}}^{(c)}(M,\cL;k)}
\def\zx{Z(\Xgn;k)}
\def\zsx{Z_s(\Xgn;k)}
\def\zasl{Z_{\{\a\}}(\Sigma_g\times S^1,\cL;k)}
\def\zt{Z_{\{\theta\}}(\Sigma_g;a)}
\def\zred{Z_{\{\xn\};\xm}^{\rm (red.)}}
\def\zirr{Z_{\{\xnp\};l}^{\rm (irr.)}}
\def\zirsp{Z_{\{\xnp\};l}^{\rm (irr. sp.)}}
\def\zgx{Z_{\{\gamma\};\gamma}(\Xgn,\cL;k)}
\def\ztirr{Z^{\rm (irr.)}_{\{\theta\}}(\Sigma_g;a)}
\def\ztred{Z^{\rm (red.)}_{\{\theta\}}(\Sigma_g;a)}
\def\ztirsp{Z^{\rm (irr. sp.)}_{\{\theta\}}(\Sigma_g;a)}

\def\mred{\cM^{\rm (red.)}_{\{\xnt\};\xmt}}
\def\mirr{\cM^{\rm (irr.)}_{\{\xnt\};\tilde{l}}}
\def\mirsp{\cM^{\rm (irr. sp.)}_{\{\xnt\};\tilde{l}}}
\def\mts{\cM_{\{\theta\}}(\Sigma_g)}
\def\ms{\cM(\Sigma_g)}
\def\tms{\tilde{\cM}(\Sigma_g)}
\def\tmns{\tilde{\cM}_n(\Sigma_g)}

\def\Pexp{\mathop{{\rm Pexp}}\nolimits}
\def\PexpA#1{\Pexp \left(\oint_{#1} A_\mu dx^\mu \right)}
\def\pjn{\prod_{j=1}^n}
\def\sjn{\sum_{j=1}^n}
\def\Tub{\mathop{{\rm Tub}}\nolimits}
\def\p{^{\prime}}
\def\tU{\tilde{U}}
\def\ffr{\phi_{\rm fr}}
\def\sign#1{\mathop{{\rm sign}\left(#1\right)}\nolimits}
\def\Nz{N_{\rm zero}}
\def\Nph{N_{\rm ph}}
\def\Xgn{X_{g,\left\{\frac{p}{q}\right\}}}
\def\tS{\tilde{S}}
\def\nga{N^g_{\{\a\}}}

\def\xn{m}
\def\xnh{\hat{\xn}}
\def\xnp{\xn\p}
\def\xnt{\tilde{\xn}}
\def\txnp{\tilde{\xn}\p}
\def\xm{m_0}
\def\xmt{\tilde{m}_0}
\def\xl{l\p}
\def\xxm{m}
\def\xxl{l\p}
\def\ttheta{\tilde{\theta}}
\def\ta{\tilde{\a}}

\def\snp{\sum_{\xn_j=0}^{p_j - 1}}
\def\snph{\sum_{\xnh_j=0}^{p_j - 1}}
\def\smj{\sum_{\{\mu\}=\pm 1}}
\def\xe{\exp\left(2\pi i \sjn \frac{r_j}{p_j}\left(K\xn_j^2 +
	\mu_j\xn_j \right)\right)}
\def\e1h{\exp\left[2\pi i \sjn \frac{r_j}{p_j}\left(K\xnh_j^2 +
	\xnh_j \right)\right]}

\def\pmj{\left(\pjn\mu_j\right)}
\def\Fbnm{F(\b;\xm,\{\xn\},\{\mu\})}
\def\fnp{\frac{\xn_j + \frac{\mu_j}{2K}}{p_j}}
\def\bst{\b_{\rm st}}
\def\Csd{C_{\rm sd}}
\def\Res{\mathop{{\rm Res}}\nolimits}
\def\Lst{\Lambda_{\rm st}}
\def\Lp{\Lambda_{\rm p}}
\def\zsp{Z_{\rm s, polar}}
\def\zstph{Z_{\rm s, st. ph.}}
\def\sl{\sum_{l=0,1}}
\def\Sym#1{\mathop{{\rm Sym}_{#1}}\nolimits}
\def\Szpm#1{\Sym{\ZZ_\pm}\left( #1 \right)}
\def\Gg{\mathop{\Gamma}\nolimits}
\def\G#1{\Gg\left(#1\right)}
\def\snpj{\sum_{0\leq \xnp_j <p_j \atop \xnp_j\in\ZZ +
	{1\over 2}q_j l}}
\def\xep{\exp \left [ 2\pi iK \sjn
	\left( {r_j\over p_j} {\xnp_j}^{2} -
	{1\over 4} s_j q_j l^2 \right)\right] }
\def\txep{\exp \left [ 2\pi iK \sjn
	\left( {r_j\over p_j} \tilde{\xn}_j^{\prime 2} -
	{1\over 4} s_j q_j l^2 \right)\right] }

\def\yep{\exp \left[ 2\pi i\sjn
	\mu_j \left( {r_j\over p_j}\xnp_j
	-{1\over 2} s_j l \right) \right] }
\def\tyep{\exp \left[ 2\pi i\sjn
	\mu_j \left( {r_j\over p_j}\txnp_j
	-{1\over 2} s_j l \right) \right] }

\def\Fbnmp{F(\b\p;\xm,\xnp_j,\mu_j)}
\def\pmu{\left( \pjn \mu_j \right)}
\def\smpj{\sum_{\{\mu\p\}=\pm 1}}
\def\snpjh{\sum_{0\leq \xnp_j\leq{p_j\over 2} \atop
	\xnp_j \in \ZZ + {1\over 2}q_j l} \frac{1}
	{\pjn \Szpm{\xnp_j\over p_j}}}

\def\lxi{\lim_{\xi\rightarrow 0^+}}
\def\smunp{\sjn{\mu\p_j \xnp_j\over p_j}}
\def\zep{\exp \left[ -{i\pi\over 2K}{H\over P}{\b\p}^2 +
	2\pi i\b\p \left( \xm - \smunp \right) \right]}
\def\denxi{\sin^{n+2g-2} \left[ {\pi\over K}(\b\p - i\xi) \right] }
\def\spm#1{\Sym{\pm}\left(#1\right)}
\def\bpp{\b^{\prime\prime}}
\def\rpnsl{\left( {r_j\over p_j}\xnp_j - {1\over 2} s_j l \right)}

\def\zspr{Z_{\rm s, polar, reg.}}
\def\zsps{Z_{\rm s, polar, sing.}}
\def\zeb{\exp \left( -{i\pi\over 2K} {H\over P} \b^2 \right) }
\def\den{\sin^{n+2g-2} \left({\pi\over K} \b \right) }
\def\zebl{\exp \left[ -{i\pi\over 2K}
{H\over P} \b^2 + 2 \pi i\b
	\sign{\smunp}
	\left(\xm - \left|\smunp\right|\right)\right] }
\def\zspec{Z_{\rm spec.}}

\def\snrp{\sin \left[ 2\pi \left( {r_j\over p_j}\xnp_j -
	{\b\p\over 2K}{\mu\p_j\over p_j}
	- {1\over 2} s_j l \right) \right] }
\def\ecass{\exp \left[ {i\pi\over 2K} \left( {H\over P} -12\sjn
	s(q_j,p_j) - 3\sign{H\over P} \right)\right] }
\def\ebst{\exp \left[ 2\pi i K \left( \sjn {r_j\over p_j} n_j^2 +
	{1\over 4}{H\over P}
	\left({\bst\over K}\right)^2\right)\right] }
\def\denphi{\sin^{n+2g-2}(2\pi\phi) }
\def\eis{e^{i{3\over 4}\pi\sign{H\over P}}}
\def\hola{\mathop{{\rm Hol}_A}\nolimits}
\def\holap{{\rm Hol}_{A\p}}
\def\mod{\mathop{{\rm mod}}\nolimits}
\def\ajst{\a_j^{\rm (st)}}
\def\denphi{\sin^{n+2g-2}(2\pi\phi)}
\def\smua{\sjn \mu_j\a_j}
\def\fri{\left({i\over 2}\right)^{n-1}}
\def\smut{\sjn\mu_j\theta_j}
\def\Td{\mathop{{\rm Td}}\nolimits}
\def\ahat{\mathop{\hat{\rm A}}\nolimits}
\def\denp{\sin^{n+2g-2}(\pi\phi)}
\def\zeph{\exp\left( - {i\pi K\over 2}
{H\over P} \phi^2\right) }
\def\fnpp{\left(\frac{\xnp_j + {\mu_j\over 2K}}{p_j} \right) }
\def\pmup{\left(\pjn \mu_j\p\right)}
\def\zephl{\exp\left[ -{i\pi K\over 2}
	{H\over P} \phi^2
	+ 2\pi iK\phi\sign{\smunp}
	\left(\xm-\left|\sjn{\mu\p_j\over p_j}
	\left(\xnp_j+{\mu_j\over 2K}\right)\right|\right)\right]}
\def\denb{\sin^{n+2g-2}\left({\pi\over K}(\b - i\xi)\right)}
\def\tK{\mathop{\tilde{K}}\nolimits}
\def\tKt{\tK^{\rm (tot)}_{n+2g-2}(\gamma;\xnp_j,\mu_j)}
\def\sjnmu{\sjn{\mu\p_j\over p_j}\left(K\xnp_j + {\mu_j\over
	2}\right)}
\def\sjnmuk{\sjn{\mu\p_j\over p_j}\left(\xnp_j + {\mu_j\over
	2K}\right)}
\def\gbr{\gamma_{\rm br}}
\def\gbms#1{G^{#1}\left( \b\p;\mu_j\right)}
\def\gbm{G\left( \b\p;\mu_j\right)}

\def\sms{\sum_{0\leq\xm \leq\left|\smunp\right|}
	\frac{\sign{\smunp}}{\spm{\xm}\spm{\xm -
	\left|\smunp\right|}} }
\def\smsold{\sum_{0\leq\xm <\left|\smunp\right|}
	\frac{\sign{\smunp}}{\spm{\xm}} }

\begin{document}

\begin{titlepage}
\centerline{\hfill                 UMTG-179-94}
\centerline{\hfill                 hep-th/9412075}
\vfill
\begin{center}

{\large \bf
Residue Formulas for the Large $k$ Asymptotics of
Witten's Invariants of Seifert Manifolds. The Case of $SU(2)$.
} \\

\bigskip
\centerline{L. Rozansky\footnote{Work supported
by the National Science Foundation
under Grant No. PHY-92 09978.
}}

\centerline{\em Physics Department, University of Miami
}
\centerline{\em P. O. Box 248046, Coral Gables, FL 33124, U.S.A.}

\vfill
{\bf Abstract}

\end{center}
\begin{quotation}

We derive the large $k$ asymptotics of the surgery formula for
$SU(2)$ Witten's invariants of general Seifert manifolds. The
contributions of connected components of the moduli space of flat
connections are identified. The contributions of irreducible
connections are presented in the residue form. This allows us to
express them in terms of intersection numbers on their moduli spaces.

\end{quotation}
\vfill
\end{titlepage}

\pagebreak
\nsection{Introduction}

Let $A_\mu$ be a connection on an $SU(2)$ bundle $E$ over a
3-dimensional manifold $M$. The Chern-Simons action is a functional
of this connection:
\qq
S_{\rm CS}=\frac{1}{2}\Tr\epsilon^{\mu\nu\rho}\int_M
d^3 x(A_\mu\partial_\nu A_\rho + \frac{2}{3}A_\mu A_\nu A_\rho),
\label{1.1}
\qqq
here $\Tr$ denotes a trace in the fundamental representation of
$SU(2)$.

Consider an $n$-component link $\cL$ in $M$. Let us attach
$\a$-dimensonal irreducible representations $V_{\a_j}$ to the
components $\cL_j$ of $\cL$. A partition function of the quantum
Chern-Simons theory with the Planck constant
\qq
\hbar=\frac{2\pi}{k},\qquad k\in \ZZ
\label{1.01}
\qqq
can be presented as a path integral taken with an appropriate measure
over the gauge equivalence classes of $A_\mu$:
\qq
\zaml=\int[\cD A_\mu] e^{\frac{i}{\hbar}S_{\rm CS}[A_\mu]}
\pjn \Tr_{\a_j} \PexpA{\cL_j},
\label{1.2}
\qqq
here $\PexpA{\cL_j} \in SU(2)$ is a holonomy of $A_\mu$ along the
contour $\cL_j$ and $\Tr_\a$ is the trace in the $\a$-dimensional
representation $V_\a$. We also use the following general notation:
${x}$ denotes a set of $n$ numbers $x_1, \ldots, x_n$. E.~Witten
noticed in~\cite{Wi1} that the partition function~(\ref{1.2}) is a
topological invariant of the (framed) manifold $M$ and link $\cL$. He
also showed that the ratio
\qq
J_{\{\a\}}(\cL;k)=\frac{Z_{\{\a\}}(S^3,\cL;k)}{\sqrt{\frac{2}{K}}
\sin \left(\frac{\pi}{K}\right) },
\qquad K=k+2
\label{1.3}
\qqq
is equal to the Jones polynomial for $q=e^{\frac{\pi i}{K}}$.

Another important result of~\cite{Wi1} is that $\zaml$ can be exactly
calculated through the surgery formula. Let us first define a
rational $(p,q)$ surgery on a knot $\cK$ belonging to a manifold $M$.
We choose a pair of cycles $C_1, C_2$ on the boundary of the tubular
neighborhood $\Tub(\cK)$. $C_1$ is a meridian, it is contractible
through $\Tub(\cK)$. $C_2$ is a parallel, it is defined by a
condition that it has a unit intersection number with $C_1$. The
parallel $C_2$ is defined only modulo $C_1$. The $(p,q)$ surgery on
$\cK$ is produced by cutting $\Tub(\cK)$ out of $M$ and then gluing
in back in such a way that the cycles $C_1$ and $C_2$ on $\partial
\Tub(\cK)$ are identified with $C_1\p = pC_1+qC_2$ and
$C_2\p=rC_1+sC_2$ on $\partial (M\setminus \Tub(\cK))$. The numbers
$r,s\in \ZZ$ are defined modulo $p,q$ by a condition
\qq
ps-qr = 1,
\label{1.4}
\qqq
which follows from the fact that $C_1\p$ and $C_2\p$ should also have
a unit intersection number. The topological class of the new manifold
$M\p$ constructed by the $(p,q)$ surgery does not depend on a
particular choice of $r$ and $s$.

Let $M\p$ be a manifold produced by rational $(p_j,q_j)$ surgeries on
the first $m$ components of the link $\cL$ in $M$. $M\p$ still
contains a link $\cL\p$ consisting of the remaining components
$\cL_{m+1},\ldots,\cL_n$ of $\cL$. According to~\cite{Wi1}, the
invariant of the new pair $M\p,\cL\p$ can be expressed in terms of
the old one through the surgery formula
\qq
Z_{\a_{m+1},\ldots,\a_n}(M\p,\cL\p;k)=
e^{i\ffr} \sum_{1\leq \a_1,\ldots,\a_m\leq K-1}
Z_{\a_1,\ldots,\a_n}(M,\cL;k) \prod_{j=1}^m \tU^{(p,q)}_{\a_j 1},
\label{1.5}
\qqq
here $\ffr$ is a framing correction phase and the matrices
$\tU^{(p,q)}_{\a\b}$ generate a $K-1$-dimensional representation of
the surgery matrices
\qq
\tU^{(p,q)}=
\left(
\matrix{p&q\cr q&s\cr}
\right)
\in SL(2,\ZZ).
\label{1.6}
\qqq
The formula for $\tU^{(p,q)}_{\a\b}$ was derived by L.~Jeffrey
in~\cite{Je1}:
\qq
\tU^{(p,q)}_{\a\b} & = & i\frac{\sign{q}}{\sqrt{2K|q|}}
e^{-\frac{i\pi}{4}\Phi(U^{(p,q)})}
\label{1.7}\\
&&\qquad\times
\sum_{\mu=\pm 1} \sum_{n=0}^{q-1} \mu \exp \left[
\frac{i\pi}{2Kq}(p\a^2 - 2\a(2Kn+\mu\b) + s(2Kn+\mu\b)^2)\right],
\nonumber
\qqq
here $\Phi(U^{(p,q)})$ is the Rademacher function defined as follows:
\qq
\Phi
\left[\matrix{p&r\cr q&s\cr}\right] =
\frac{p+s}{q} - 12 s(p,q),
\label{1.8}
\qqq
$s(p,q)$ is a Dedekind sum
\qq
s(p,q) = \frac{1}{4q} \sum_{j=1}^{|q|-1}
\cot\left(\pi\frac{j}{q}\right)
\cot\left(\pi\frac{pj}{q}\right).
\label{1.9}
\qqq

N.~Reshetikhin and V.~Turaev showed in~\cite{ReTu}
that the surgery formula~(\ref{1.5}) defines a
topological invariant, without relying on the path integral
representation~(\ref{1.2}). This made the whole theory mathematically
rigorous. They also formulated a set of general conditions on the
components of eq.~(\ref{1.5}) for it to define an invariant. The
problem with the surgery formula~(\ref{1.5}) is however that it does
not obviate the relation between the ``quantum'' invariant $\zaml$
and the well-known classical invariants of $M$ and $\cL$ such as
Betti numbers, linking nubmers, \etc. A possible remedy is to study
the large $k$ asymptotic behavior of $\zaml$ by applying a stationary
phase approximation to the path integral~(\ref{1.2}). The stationary
points of the phase~(\ref{1.1}) are $SU(2)$ flat connections on $M$.
Let $\cM$ be their moduli space, $\cM_c$ being its connected
components numbered by the index $c$. Each component $\cM_c$ gives
its own contribution $\zcaml$ to the total invariant:
\qq
\zaml=\sum_c \zcaml.
\label{1.10}
\qqq
The individual contributions are presented as asymptotic series in
$\hbar$ (or the exponentials thereof):
\qq
\zcaml = (2\pi\hbar)^{\frac{\Nz}{2}}
\exp \left(\frac{i}{\hbar}S_{\rm CS}^{(c)}\right)
\left[ \sum_{n=1}^\infty \hbar^{n-1}\Delta_n^{(c)} \right],
\label{1.11}
\qqq
or, equivalently,
\qq
\zcaml = (2\pi\hbar)^{\frac{\Nz}{2}}
\exp \left[\frac{i}{\hbar}
\left(
S_{\rm CS}^{(c)}
+ \sum_{n=1}^\infty S_n^{(c)} \hbar^n
\right)\right].
\label{1.12}
\qqq
Here $S_{\rm CS}^{(c)}$ is a Chern-Simons action of connections of
$\cM_c$ and
\qq
\Nz = \dim H_c^0 - \dim H_c^1,
\label{1.13}
\qqq
$H_c^{0,1}$ being the cohomologies of the covariant (with respect to
$A_\mu$) derivative $D$. The coefficients $\Delta_n^{(c)}, S_n^{(c)}$
are called $n$-loop corrections. The expression for the 1-loop
correction was derived in~\cite{Wi1}, ~\cite{FrGo} and~\cite{Je1}
(some details were added in~\cite{Ro1}):
\qq
\Delta_1^{(c)} \equiv e^{i S_1^{(c)}} & = & \frac{1}{\Vol(H_c)}
\exp\left[\frac{i}{\pi}\left(S_{\rm CS}^{(c)} - \frac{i\pi}{8} \Nph
\right)\right]
\label{1.14}\\
&&\quad\times
\int_{\cM_c}\sqrt{|\tau_R|} \pjn \Tr_{a_j} \PexpA{\cL_j}.
\nonumber
\qqq
In this formula $H_c$ is an isotropy group of $\cM_c$ (\ie a subgroup
of $SU(2)$ which commutes with the holonomies of connections of
$\cM_c$), $\Nph$ is expressed in~\cite{FrGo} as
\qq
\Nph = 2I_c + \dim H_c^0 + \dim H_c^1 + 3(1+b^1_M),
\label{1.15}
\qqq
here $I_c$ is a spectral flow of the operator $L_-=\star D + D \star$
acting on 1- and 3-forms on $M$, $b^1_M$ is the first Betti number of
$M$. $\tau_R$ is the Reidemeister-Ray-Singer torsion. L.~Jeffrey
observed in~\cite{Je1} that $\sqrt{|\tau_R|}$ defines a ratio of
volume forms on $\cM_c$ and $H_c$.

The higher loop corrections $\Delta_n^{(c)}, S_n^{(c)}$ come from the
$n$-loop Feynman diagrams. They are expressed as multiple integrals
of the products of propagators taken over the manifold $M$ and the
link $\cL$ (see, \eg~\cite{AxSi}, ~\cite{BN1} and references therein
for details).

The asymptotic formulas~(\ref{1.10})-(\ref{1.12}) follow from the
path integral of eq.~(\ref{1.3}) and can not be derived directly (at
least, at this point) from the surgery formula~(\ref{1.5}). In other
words, the asymptotic properties of the \rhs of eq.~(\ref{1.5}) are
not immediately obvious. Therefore it is interesting to take the
surgery formula for the invariant of a particular simple manifold and
try to find its large $k$ asymptotics in order to compare it with
eq.~(\ref{1.14}) and multiloop Feynman diagrams. This program was
initiated by D.~Freed and R.~Gompf in~\cite{FrGo}. They observed a
numerical correspondence between the invariants of some lens
spaces and Seifert manifolds calculated through surgery formula and
the predictions of eqs.~(\ref{1.10}),~(\ref{1.14}) for large values
of $k$. L.~Jeffrey worked out the full asymptotic expansion of the
invariants of lens spaces as well as some mapping class tori
in~\cite{Je1}. She checked analyticly that the classical and 1-loop
parts of the flat connection contributions were equal to the
Chern-Simons action and the \rhs of eq.~(\ref{1.14}).

In our previous paper~\cite{Ro1} we studied the large $k$ asymptotics
of the invariant of Seifert manifolds constructed by rational
surgeries on the fibers of $S^2\times S^1$. We demonstrated the
consistency between our results and eqs.~(\ref{1.10}),~(\ref{1.14})
for the case of 3-fibered spaces. We also found that the
contributions of irreducible flat connections were finite loop exact.
This means that (up to minor details) the asymptotic series
$\sum_{n=1}^\infty\Delta_n^{(c)}\hbar^{n-1}$ of eq.~(\ref{1.11})
appeared to be finite polynomials for the case when $\dim H_c=0$.
Such behavior is similar to the one observed in~\cite{Wi3} for the 2d
Yang-Mills theories and explained there by a non-abelian
localization.

In this paper we study the large $k$ asymptotics of $SU(2)$ Witten's
invariant of general Seifert manifolds $\Xgn$. We calculate all
contributions $Z^{(c)}(\Xgn;k)$ (Proposition~\ref{p3.1.1})
and relate
them to connected components of the moduli space of flat connections
(Proposition~\ref{p4.2.1}).
Our formulas express the contributions of
irreducible connections as residues, which makes them look similar to
the non-abelian localization formulas of~\cite{JeKi} and~\cite{Sz2}.
By comparing our expressions with the residue formulas for
intersection numbers derived in~\cite{JeKi} and conjectured
in~\cite{Sz2} we express the contributions of irreducible connections
in terms of intersection numbers on their moduli spaces
(Proposition~\ref{p5.1.3}). As a byproduct of our calculations we
derive the full asymptotic expansion of the partition function of 2d
$SU(2)$ Yang-Mills theory on a Riemann surface with punctures,
including the contributions of constant curvature reducible
connections (Proposition~\ref{p5.1.2}). In Appendix~\ref{a1} we
discuss the alternative way of deriving the asymptotics of Witten's
invariants of Seifert manifolds which relates them to Kostant's
partition function (this is analogous to the relation between the
intersection numbers and Duistermaat-Heckman polynomial discussed
in~\cite{JeKi}). In Appendix~\ref{a2} we use the moduli space of
twisted flat $SU(2)$ connections in order to get rid of
singularities of the moduli space of untwisted connections and to
simplify some residue and intersection number formulas.

\nsection{A Surgery Formula for Seifert Manifolds}
\label{s2}

The simplest way to construct a Seifert manifold $\Xgn$ is to perform
$n$ rational surgeries on the manifold $\Sigma_g\times S^1$,
$\Sigma_g$ being a $g$-handled Riemann surface. Choose $n$ points
$P_j$, $1\leq j\leq n$ on $\Sigma_g$ and consider an $n$-component
link $\cL$ in $\Sigma_g\times S^1$ formed by the loops $P_j\times
S^1$. The Seifert manifold $\Xgn$ is constructed by $n$
rational $(p_j,q_j)$
surgeries on the link components $\cL_j$. The surgery
formula~(\ref{1.5}) tells us that
\qq
\zx=e^{i\ffr} \sum_{1\leq \{\a\}\leq K} \zasl \pjn
\tU^{(p_j,q_j)}_{\a_j}.
\label{2.1.1}
\qqq
The framing correction $\ffr$ for this surgery was calculated
in~\cite{FrGo}:
\qq
\ffr = \frac{\pi}{4}\left(1-\frac{2}{K}\right)
\left[\sjn \Phi(U^{(-q_j,p_j)}) + 3\sign{\frac{H}{P}}\right],
\label{2.1.2}
\qqq
here we used a notation
\qq
P=\pjn p_j,\qquad H=P\sjn \frac{q_j}{p_j}.
\label{2.1.3}
\qqq
The invariant $\zasl$ is equal to the Verlinde number, \ie to the
number of conformal blocks of the $SU(2)$ WZW theory at level $k$ for
the surface $\Sigma_g$ with $n$ insertions of the primary fields
$\cO_{\a_j}$
which correspond to the representations $V_{\a_j}$. The number $\nga$
is given by the Verlinde formula
\qq
\zasl = \nga = \sum_{\b=1}^{K-1}\frac{\pjn \tS_{\a_j\b}}
{\tS_{\b 1}^{n+2g-2}},
\label{2.1.4}
\qqq
here $S$ is an $SL(2,\ZZ)$ matrix which interchanges a parallel and a
meridian:
\qq
S=\left(\matrix{0&-1 \cr 1&0 \cr}\right) \in SL(2,\ZZ)
\label{2.1.5}
\qqq
and $\tS_{\a\b}$ is its $K-1$-dimensional representation:
\qq
\tS_{\a\b}= \sqrt{\frac{2}{K}}\sin\left(\frac{\pi}{K}\a\b\right).
\label{2.1.6}
\qqq
By substituting eqs.~(\ref{2.1.2}) and~(\ref{2.1.4}) into
eq.~(\ref{2.1.1}) and using an obvious relation
$SU^{(p,q)}=U^{(-q,p)}$ we arrived at the following equation:
\qq
\zx&=&e^{i\ffr} \sum_{\b=1}^{K-1} \frac{\pjn \tU^{(-q_j,p_j)}}
{\tS_{\b1}^{n+2g-2}}
\label{2.1.7}\\
&=&\frac{i^n K^{g-1}}{2^{n+g-1}} \frac{\sign{P}}{\sqrt{|P|}}
e^{\frac{3}{4}i\pi\sign{\frac{H}{P}}}
\nonumber\\
&&\qquad\times
\ecass
\zsx,
\nonumber\\
\zsx &=& \sum_{\b=1}^{K-1} \frac {\exp \left( - \frac{i\pi}{2K}
\frac{H}{P} \b^2 \right) } { \sin^{n+2g-2} \left( \frac{\pi}{K} \b
\right) }
\snp \smj \pmj \xe
\nonumber\\
&&\quad\times
\exp \left( -\frac{i\pi}{K}\b \sjn \frac{2K\xn_j +
\mu_j} {p_j} \right).
\label{2.1.8}
\qqq
Here we split the invariant $\zx$ into a product of lengthy numerical
factors and a sum $\zsx$ whose large $k$ asymptotics has to be
determined. Note that this sum takes a slightly different form if we
substitute $\xnh_j = \mu_j \xn_j$ for $\xn_j$:
\qq
\zsx & = & (-2i)^n \snph \e1h
\label{2.1.9}\\
&&\qquad\times
\sum_{\b=1}^{K-1}
\frac {\exp \left( - \frac{i\pi}{2K}
\frac{H}{P} \b^2 \right) } { \sin^{n+2g-2} \left( \frac{\pi}{K} \b
\right) }
\pjn \sin \left( 2\pi\b \frac {\xnh_j + \frac{1}{2K}}{p_j}
\right).
\nonumber
\qqq
This expression bears a close resemblance to the following two
objects: Verlinde numbers and a partition function of the 2d
Yang-Mills theory. With the substitution of eq.~(\ref{2.1.6}),
Verlinde formula~~(\ref{2.1.4}) turns into
\qq
\nga = \left( \frac{K}{2} \right)^{g-1} \sum_{\b=1}^{K-1}
\frac {\pjn \sin \left( \frac{\pi}{K}\b\a_j \right) }
{\sin^{n+2g-2} \left( \frac{\pi}{K}\b \right) }.
\label{2.1.10}
\qqq
According to~\cite{Wi2}, a partition function of a 2d Yang-Mills
theory with the coupling constant $a$ defined on a unit area surface
$\Sigma_g$, which has $n$ punctures with the holonomies
\qq
\PexpA{} = \exp (2\pi i \sigma_3 \theta_j), \qquad
\sigma_3 = \left( \matrix{1&0\cr 0&-1\cr} \right)
\label{2.1.11}
\qqq
around them, is equal to
\qq
\zt = \frac {1} { 2^{g-1} \pi^{n+2g-2} }
\sum_{\b\geq 1} \frac { e^{-a\b^2} } { \b^{n+2g-2} }
\pjn \sin (2\pi\b\theta_j).
\label{2.1.12}
\label{2.1.13}
\qqq

The similarity between the sums in eqs.~(\ref{2.1.9}),~(\ref{2.1.10})
and~(\ref{2.1.12}) becomes apparent if we put
\qq
\a_j = 2K \frac {\xnh_j + \frac{1}{2K}} {p_j},\qquad
\theta_j = \frac {\xnh_j + \frac{1}{2K}} {p_j},\qquad
a = \frac{i\pi}{K} \frac{H}{P}.
\label{2.1.14}
\qqq
The sum~(\ref{2.1.9}) is a generalization of the other two sums: it
has a quadratic exponent of eq.~(\ref{2.1.12}) and a sine in
denominator of eq.~(\ref{2.1.10}). The difference between the ranges
of summation in the sums~(\ref{2.1.9}) and~(\ref{2.1.11}) does not
affect the similarity of calculation of their asymptotics as we will
see in the next section. Note however that we cannot multiply the
summand of eq.~(\ref{2.1.10}) by an arbitrary quadratic exponential.
The exponent of eq.~(\ref{2.1.9}) is special:
the exponential is periodic in $\b$
after the sum over $\xn_j$ is taken.

\nsection{A Residue Calculation of Asymptotics}
\label{s3}
Now we turn directly to the asymptotic calculation of the
sum~(\ref{2.1.8}). We convert it into a sum over $\b\in\ZZ$ in two
steps. By slightly shifting the argument of the denominator along the
imaginary axis we can double the range of summation:
\qq
\zsx & = & \snp \exp \left( 2\pi i K \sjn \frac{r_j}{p_j}\xn_j^2
\right)\, \lim_{\xi \rightarrow 0^+} \frac{1}{2} \sum_{\b=-K+1}^{K}
\frac {\exp \left( -\frac{i\pi}{2K} \frac{H}{P} \b^2 \right)}
{\sin^{n+2g-2}\left(\frac{\pi}{K}(\b-i\xi)\right)}
\label{3.1.1}\\
&&\qquad\times
\smj \left\{ \pjn \mu_j \exp \left[ \frac {2\pi i}{p_j}
\left(r_j \mu_j \xn_j - \b(\xn_j +
\frac{\mu_j}{2K}) \right) \right] \right\}
\nonumber
\qqq
Indeed, the product of sines kills the summand at $\b=0$ (to see that
the same happens at $\b=K$ combine the terms at $\xn_j$ and $q_j -
\xn_j$). If the product of sines is absent (as it happens for the
sums~(\ref{2.1.10}) and~(\ref{2.1.12}) if $n=0$) we may add an extra
factor
\qq
\frac {\sin \left( \frac{\pi}{K}\b\right)}
{\sin \left[ \frac{\pi}{K}(\b - i\xi)\right] }
\label{3.1.2}
\qqq
that will take care of $\b=0,K$.

As the next step, we extend the sum over $\b$ to all integer numbers
by using the following simple lemma:
\begin{lemma}
\label{l3.1.1}

If the function $f(\b)$ defined on $\ZZ$ has a period $T$ then
\qq
\sum_{\b=0}^{T-1} f(\b) = \frac{1}{T} \lim_{\epsilon\rightarrow 0}
\sqrt{\epsilon} \sum_{\b \in \ZZ} f(\b) e^{-\pi \epsilon \b^2}.
\label{3.1.3}
\qqq
\end{lemma}
As a result,
\qq
\zsx & = &
\frac{1}{2K}\lim_{\epsilon\rightarrow 0} \sqrt{\epsilon}
\snp \exp \left( 2\pi i K \sjn \frac{r_j}{p_j}\xn_j^2
\right)\, \lim_{\xi \rightarrow 0^+} \frac{1}{2}
\sum_{\b \in \ZZ}
e^{-\pi\epsilon\b^2}
\label{3.1.4}\\
&\!\!\times&
\frac {\exp \left( -\frac{i\pi}{2K} \frac{H}{P} \b^2 \right)}
{\sin^{n+2g-2}\left(\frac{\pi}{K}(\b-i\xi)\right)}
\!\!\smj \left\{ \pjn \mu_j \exp \left[ \frac {2\pi i}{p_j}
\left(r_j \mu_j \xn_j - \b(\xn_j +
\frac{\mu_j}{2K}) \right) \right] \right\}
\nonumber
\qqq
Thus we eliminated the difference in the summation range between
eqs.~(\ref{2.1.9}), ~(\ref{2.1.10}) and eq.~(\ref{2.1.12}).

At this point we can use the Poisson resummation formula
\qq
\sum_{\b \in \ZZ} f(\b) = \sum_{m \in \ZZ}
\int_{-\infty}^{+\infty}
d\b
\exp (2 \pi i m \b) f(\b),
\label{3.1.5}
\qqq
which tells us that
\qq
\lefteqn{
\zsx  =  \frac{1}{2K} \lim_{\epsilon\rightarrow 0} \sqrt{\epsilon}
\snp \sum_{\xm\in\ZZ} \smj \pmj
}
\label{3.1.6}\\
&&\qquad\times
\xe
\lim_{\epsilon\rightarrow 0} \frac{1}{2} \int_{-\infty}^{+\infty}
d\b e^{-\pi\epsilon\b^2} \Fbnm,
\nonumber
\qqq
\qq
\Fbnm  =  \frac { \exp \left[ -\frac{i\pi}{2K} \frac{H}{P} \b^2
+ 2 \pi i \b \left( \xm - \sjn \fnp \right) \right] }
{\sin^{n+2g-2} \left[ \frac{\pi}{K}(\b - i\xi)\right] }.
\label{3.1.7}
\qqq

A substituition $\b=K\tilde{\b}$ in the integral~(\ref{3.1.6}) would
demonstrate explicitly the applicability of the stationary phase
approximation in the limit $K\rightarrow \infty$. The stationary
phase point for the phase of the integrand~(\ref{3.1.7}) is
\qq
\bst = 2K \frac{P}{H} \left( \xm - \sjn \frac{\xn_j}{p_j}\right).
\label{3.1.8}
\qqq
The steepest descent contour $\Csd(\bst)$ in the complex $\b$ plane
is the line
\qq
\Im \b = - \sign{\frac{H}{P}} (\Re \b - \bst ).
\label{3.1.10}
\qqq

In the process of being deformed from its original form $\Im \b =0$
to~(\ref{3.1.10}) the integration contour crosses those poles
\qq
\b_l = K (l + i\xi)
\label{3.1.11}
\qqq
of the integrand~(\ref{3.1.7}) for which
\qq
\sign{\frac{H}{P}} (\bst - Kl) > 0.
\label{3.1.12}
\qqq
Therefore to the leading order in $\epsilon$
\qq
\int_{-\infty}^{+\infty} d\b e^{-\pi\epsilon\b^2} \Fbnm  & = &
e^{-\pi\epsilon\bst^2} \int_{\Csd(\bst)} d\b \Fbnm
\label{3.1.13}\\
&& \qquad
+ 2\pi i
\sum_{l\in\ZZ \atop \sign{\frac{H}{P}} (\bst - Kl) > 0}
\Res_{\b=\b_l} \Fbnm.
\nonumber
\qqq

Let us substitute this expression into eq.~(\ref{3.1.6}) and take the
sum over $\xn_j,\mu_j$ and $\xm$. The function $\zsx$ will be
presented as a sum of the contributions of all stationary phase
points~(\ref{3.1.8}) with $\xm$ and $\xn_j$ belonging to the
summation range of eq.~(\ref{3.1.6}) as well as the contributions of
the poles~(\ref{3.1.11}). Both stationary points and poles form
1-dimensional lattices $\Lst$ and $\Lp$, which are invariant under
the shift
\qq
\b \rightarrow \b + 2K
\label{3.1.14}
\qqq
and (if we put $\xi = 0$ in $\Lp$) a reflection
\qq
\b \rightarrow -\b
\label{3.1.15}
\qqq
The function
\qq
\xe \Fbnm
\label{3.1.16}
\qqq
is invariant under the same transformations in the limit $\xi
\rightarrow 0$ if we combine the shift~(\ref{3.1.14}) with the shift
of $\xn_j$
\qq
\xn_j \rightarrow \xn_j - q_j,\qquad 1\leq j\leq n
\label{3.1.17}
\qqq
and the reflection~(\ref{3.1.15}) with the reflections
\qq
\xm \rightarrow -\xm,\qquad
\xn_j \rightarrow -\xn_j,\qquad
\mu_j \rightarrow -\mu_j,\qquad
1\leq j\leq n.
\label{3.1.18}
\qqq
An extra symmetry
\qq
\xm \rightarrow \xm - 1,\qquad
\xn_j \rightarrow \xn_j + p_j
\label{3.1.19}
\qqq
helps us to keep
$\xn_j$ within their summation range. Thus we conclude
that the contributions of the stationary points $\Lst$ and poles
$\Lp$ have the symmetries~(\ref{3.1.14}) and~(\ref{3.1.15}). Now we
can apply the Lemma~\ref{l3.1.1} ``backwards'' to the contributions
of $\Lst$ and $\Lp$. We remove $\frac{1}{2K}
\lim_{\epsilon\rightarrow 0}\sqrt{\epsilon}$ from eq.~(\ref{3.1.6})
while taking only the contributions of the poles $\b_0$ and $\b_1$
and of the stationary points $0\leq\bst\leq K$
(if $\bst\neq 0,K$, then
their contributions should be doubled in view of the
symmetry~(\ref{3.1.15})):
\qq
\zsx & = & \zsp + \zstph,
\label{3.1.20}
\qqq
\qq
\zsp & = & \pi i \sl \snp \smj
\sum_{\xm \in \ZZ \atop \sign{\frac{H}{P}}(\bst - Kl)>0}
\left( \pjn \mu_j \right) \xe
\nonumber\\
&&\qquad\times
\lim_{\xi\rightarrow 0^+} \Res_{\b=\b_l} \Fbnm.
\label{3.1.21}\\
\zstph & = & \snp \smj \sum_{\xm\in\ZZ \atop 0\leq \bst\leq K}
\frac{1}{\Szpm{\frac{\bst}{2K}}}
\left( \pjn \mu_j \right) \xe
\nonumber\\
&&\qquad\times
\lim_{\xi\rightarrow 0^+}
\int_{\Csd(\bst)}d\b \Fbnm.
\label{3.1.22}
\qqq
Here we used the following notation: let $G$ be a group acting on a
set $X$. For $x\in X$, we denote by $\Sym{G}(x)$ the number of
elements of $G$ which leave $x$ invariant. In the future we will need
two groups: the group of reflections $\pm$ (its only nontrivial
element multiplies real numbers by $-1$) and the group of ``affine''
reflections $\ZZ_\pm$ which combines reflections of $\pm$ with the
shifts by integer numbers.

If we substitute eqs.~(\ref{3.1.20})-(\ref{3.1.22}) into
eq.~(\ref{2.1.7}) we will see that the whole invariant $\zx$ turns
into a sum of polar and stationary phase contributions.
Let us first calculate the contribution of the stationary phase
points $\bst\neq 0,K$. We introduce a new integration variable
\qq
x = \frac{\b - \bst}{K} e^{-\frac{i\pi}{4}\sign{\frac{H}{P}}},
\label{3.1.23}
\qqq
so that
\qq
\lefteqn{
\lim_{\xi\rightarrow 0^+} \int_{\Csd(\bst)} d\b \Fbnm
}
\label{3.1.24}\\
& = &
K e^{\frac{i\pi}{4}\sign{\frac{H}{P}}}
\exp \left[ \frac{i\pi}{2K} \left(\frac{H}{P} \bst^2 -
2\bst \sjn \frac{\mu_j}{p_j} \right) \right]
\nonumber\\
&&\qquad\times
\int_{-\infty}^{+\infty} dx \frac
{\exp \left( -\frac{\pi K}{2} \left|\frac{H}{P}\right| x^2
- i\pi x e^{\frac{i\pi}{4}\sign{\frac{H}{P}}} \sjn \frac{\mu_j}{p_j}
\right)}
{\sin^{n+2g-2} \left(\frac{\pi}{K}\bst + \pi x
e^{\frac{i\pi}{4}\sign{\frac{H}{P}}} \right)}
\nonumber\\
&=&
2K \exp\left( \frac{i \pi}{2K} \frac{H}{P} \bst^2 \right)
\sum_{\xl=0}^{\infty}\frac{\G{\xl+{1\over 2}}}{(2\xl)!}
\left( {1 \over 2\pi i K}{P\over H}\right)^{\xl+{1 \over 2}}
\left.\partial_\phi^{(2\xl)}\frac
{e^{-2\pi i\phi\sjn{\mu_j\over p_j}}}
{\sin^{n+2g-2}(2\pi\phi)}
\right|_{\phi={\bst \over 2K}}
\nonumber
\qqq
Here we expanded the preexponential factor of the intergrand
in powers
of $x$ and integrated the series term by term.

The case of $\bst=0,K$ requires a more careful consideration because
the stationary phase point coincides with one of the
poles~(\ref{3.1.11}). First of all, we introduce new variables
\qq
\b\p = \b - Kl,\qquad \xnp_j =\xn_j + {1\over 2}q_j l,
\label{3.1.27}
\qqq
in which the contribution of $\bst=Kl$ is equal to
\qq
\zspec^{(l)} & = &
{(-1)^{nl} \over 2}
\snpj \smj
\pmu
\label{x3.1.1}\\
&&\qquad\times
\xep \yep
\nonumber\\
&&\qquad\qquad\times
\lim_{\xi\rightarrow 0^+}
\int_{C(\xi)} d\b\p\gbm.
\nonumber
\qqq
Here
\qq
\gbm = \frac
{ \left[ -{i\pi \over 2K} \left( {H\over P} \b^{\prime 2} +
2 \b\p \sjn {\mu_j \over p_j} \right) \right] }
{ \sin^{n+2g-2} \left( {\pi\over K} \b\p \right) }
\label{x3.1.2}
\qqq
and the contour $C(\xi)$ is described by equation
\qq
\Im \b\p = - \sign{ H\over P} \Re \b\p - \xi.
\label{x3.2.3}
\qqq

Let us split the function $\gbm$ into odd and even parts:
\qq
\gbms{\pm} = {1\over 2} \left( \gbm \pm G(-\b\p;\mu_j) \right).
\label{x3.1.4}
\qqq
To calculate the integral of $\gbms{-}$ we double the integration
contour and then close it:
\qq
\lxi \int_{C(\xi)} d\b\p \gbms{-} & = &
{1\over 2} \lxi \left[ \int_{C(\xi)} - \int_{C(-\xi)} \right]
d\b\p \gbms{-}
\label{x3.1.5}\\
& = &
\pi i \Res_{\b\p = 0} \gbms{-} = \pi i \Res_{\b\p = 0} \gbm.
\nonumber
\qqq
We substituted $\gbm$ for $\gbms{-}$ because $\gbms{+}$ has zero
residue.

To integrate $\gbms{+}$  we introduce a bew integration variable
\qq
x = \left( {\b\p \over K} e^{-{i\pi\over 2} \sign{H\over P} }
\right)^2,
\label{x3.1.6}
\qqq
so that the integration contour $C(\xi)$ folds into two branches: one
over and one under the positive semi-axis in the comples $x$ plane.
The expansion of the preexponential factor $\sin^{2-2g-n} \left(
{\pi\over K} \b\p \right)$ in powers of $x$ leads to
$\Gamma$-function type integrals:
\qq
\lefteqn{
\lim_{\xi\rightarrow 0^+} \int_{C(\xi)} d\b \gbms{+}
}
\label{3.1.26}\\
& = &
{2K\over (2\pi)^{n+2g-2}}
\sum_{\xl\geq 0 \atop \xl-n\in 2\ZZ}
\left({1 \over 2\pi i K}{P\over H}\right)^{\xl-n-2g+3 \over 2}
\frac{\G{\xl-n-2g+3 \over 2}}{\xl!}
\nonumber\\
&&\qquad\times
\partial_\phi^
{\xl} \left. \left[
e^{-2\pi i\phi \sjn{\mu_j\over p_j}}
\left(\frac{2\pi\phi}
{\sin(2\pi\phi)}\right)^{n+2g-2}\right]\right|_{\phi=0}.
\nonumber
\qqq
The $\Gamma$-function in this equation is well-defined even if its
argument is negative, because it is always half-integer.

It remains now
to substitute eq.~(\ref{3.1.24}) into eq.~(\ref{3.1.22})
and~(\ref{x3.1.5}),~(\ref{3.1.26}) into~(\ref{x3.1.1}). Recall
that $\zspec^{(0,1)}$ represents the contributions of $\bst=0,K$
to $\zstph$. The sum over $\xm$ in eq.~(\ref{3.1.22}) is
finite due to the condition $0\leq\bst\leq K$.

Now we turn to the polar contributions. The calculation of the
residue in eq.~(\ref{3.1.21}) is straightforward. The problems come
from the condition~(\ref{3.1.12}). Consider a contribution of a
general pole $\b_l$. We introduce the new variables~(\ref{3.1.27}),
so that the pole at $\b = \b_l$ corresponds to the pole at $\b\p =
\b\p_0 = i\xi$. In the new variables the contribution of $\b_l$ to
$\zsp$ is equal to
\qq
\zsp^{(l)} & = &
\pi i (-1)^{nl} \snpj \smj \sum_{\xm\in\ZZ \atop
\bst\p\sign{H\over P}
> 0} \pmu
\label{3.1.28}\\
&&\qquad\times
\xep \yep
\nonumber\\
&&\qquad\qquad\times
\lim_{\xi\rightarrow 0^+} \Res_{\b\p=i\xi} \Fbnmp,
\nonumber
\qqq
here
\qq
\bst\p = 2K {P\over H} \left(\xm - \sjn {\xnp_j\over p_j} \right).
\label{3.1.29}
\qqq
We used the symmetry~(\ref{3.1.19}) in order to reduce the range of
summation over $\xnp_j$ to $0\leq\xnp_j<p_j$. The numbers $\xnp_j$
are integer or half-integer depending on the parity of $q_j$ and $l$.
The same symmetry~(\ref{3.1.19}) allows us to further transform the
sum $\snpj$ into
\qq
\smpj\snpjh
\qqq
if we substitute $\mu\p_j\xnp_j$ for
$\xnp_j$ everywhere in eq.~(\ref{3.1.21}). After taking a sum over
$\{\mu\}=\pm 1$, we arrive at the following expression
\qq
\zsp^{(l)} & = & \pi i (-1)^{nl} (2i)^n \snpjh \xep
\nonumber\\
&&\times
\smpj \sum_{\xm > \smunp} \lxi
\Res_{\b\p=i\xi} \frac{\zep}{\denxi}
\nonumber\\
&&\qquad\times
\pjn \mu\p_j \sin \left[ 2\pi \left( {r_j\over p_j}\xnp_j
-{\b\p\over 2K}{\mu\p_j\over p_j} -
{1\over 2} s_j l \right) \right].
\label{3.1.30}
\qqq

We can extend the sum over $\xm$ to
\qq
\sum_{\xm\geq \smunp} {1\over \spm{\xm - \smunp}},
\qqq
if we transfer the polar contributions~(\ref{x3.1.5}) from $\zstph$
to $\zsp$. Then this sum can be split in two parts with the help of
the formula
\qq
\sum_{\xm \geq a} { f(\xm) \over \spm{\xm - a}}
& = &
\sum_{\xm \geq 0}\frac{f(\xm)}{\spm{\xm}}
\label{3.1.31}\\
&&
-
\sum_{0\leq\xm \leq |a|}
\frac{\sign{a}}{\spm{\xm}\spm{\xm - |a|}} f(\sign{a}\xm).
\nonumber
\qqq

The residue in eq.~(\ref{3.1.30}) is calculated at $\b\p = i\xi$, so
we can assume that $\Im\b\p>0$. Then the sum
$\sum_{\xm \geq 0}\frac{1}{\spm{\xm}}$ can be easily
calculated:
\qq
\sum_{\xm \geq 0}\frac{e^{2\pi i\b\p\xm}}{\spm{\xm}}
= {i\over 2} \cot(\pi\b\p).
\label{3.1.32}
\qqq
Let us introduce the variable $\bpp = \b\p - i\xi$. The residue in
$\bpp$ is calculated at $0$. A dependence on $\xi$ in the vicinity of
this point is nonsingular except for the factor
$\cot[\pi(\bpp+i\xi)]$ coming from the sum~(\ref{3.1.32}). Since the
range of summation in $\sum_{\xm\geq 0}$ does not depend on
$\mu\p_j$, the sum $\smpj$ in eq.~(\ref{3.1.30}) can be calculated
explicitly:
\qq
\lefteqn{
\smpj \mu\p_j \exp \left( -2\pi i\b\p {\mu\p_j\xnp_j\over p_j}
\right) \snrp
}
\label{3.1.33}\\
&=& -2i \sin \left[ 2\pi \rpnsl \right]
\cos \left( {\pi\over K} {\b\p \over p_j}\right)
\sin \left( 2\pi\b\p {\xnp_j\over p_j} \right)
\nonumber\\
&&
- 2\cos \left[ 2\pi \rpnsl \right]
\sin \left( {\pi\over K} {\b\p \over p_j}\right)
\cos \left( 2\pi\b\p {\xnp_j\over p_j} \right)
\nonumber
\qqq
The factors $\sin\left[ 2\pi(\bpp + i\xi) {\xnp_j\over p_j} \right]$
and $\sin \left( {\pi\over K} {\bpp + i\xi\over p_j}\right)$ cancel
the singularity of $\cot[\pi(\bpp + i\xi)]$ if $n\geq 1$. Otherwise
the singularity will be canceled by the extra factor~(\ref{3.1.2}).
As a result, we may simply put $\xi = 0$ in eq.~(\ref{3.1.30}):
\qq
\zsp^{(l)} & = & \pi i (-1)^{nl} (2i)^n \!\!\!\!\! \snpjh \xep
\nonumber\\
&&\qquad\times
(\zspr^{(l)} + \zsps^{(l)}),
\label{3.1.34}\\
\zspr^{(l)} & = & {i\over 2} (-2)^n \Res_{\b=0} \left\{
\frac{\zeb}{\den} \cot(\pi\b)
\right.\nonumber\\
&&\times
\pjn \left[
i \sin \left[ 2\pi \rpnsl \right]
\cos \left( {\pi\over K} {\b\p \over p_j}\right)
\sin \left( 2\pi\b\p {\xnp_j\over p_j} \right)
\right.\nonumber\\
&&\left.\left.\qquad
+ \cos \left[ 2\pi \rpnsl \right]
\sin \left( {\pi\over K} {\b\p \over p_j}\right)
\cos \left( 2\pi\b\p {\xnp_j\over p_j} \right)
\right]\right\},
\label{3.1.35}\\
\zsps^{(l)} & = & - \smpj \sms
\nonumber\\
&&\times
\Res_{\b=0} \frac {\zebl} {\den}
\nonumber\\
&&\qquad\times
\pjn \mu\p_j \snrp.
\label{3.1.36}
\qqq
The reason why we call the sum~(\ref{3.1.36}) singular (apart from
the apparent ugliness of the sum over $\xm$) is that it seems to be
related to a singularity in the ``underlying'' moduli space. Note
that $\zspr^{(l)}=0$ if $g=0$ because the function whose residue is
calculated in eq.~(\ref{3.1.35}) is nonsingular at $\b = 0$.

Now it just remains to combine together
eqs.~(\ref{2.1.7}),~(\ref{3.1.20}),~(\ref{3.1.22}),~(\ref{3.1.24}),
{}~(\ref{3.1.26}),~(\ref{3.1.34})-(\ref{3.1.36}) into one proposition:
\begin{proposition}
\label{p3.1.1}
The large $k$ asymptotics of Witten's invariant of a Seifert manifold
is a sum of a finite number of contributions:
\qq
\zx & = & \snp \sum_{\xm\in\ZZ \atop 0<\bst<K} \zred
+ \sl \sum_{0\leq \xnp_j \leq{p_j\over 2} \atop \xnp_j \in
\ZZ + {1\over 2} q_j l,\; \sjn {\pm\xnp_j\over p_j}\not\in\ZZ} \zirr
\nonumber\\
&&\qquad
+ \sl \ \sum_{0\leq \xnp_j \leq{p_j\over 2} \atop \xnp_j \in
\ZZ + {1\over 2} q_j l,\;\exists\mu\p_j=\pm 1:\;\smunp\in\ZZ} \zirsp,
\label{3.1.37}
\qqq
\qq
\zred  & = & (-1)^n {2\over \sqrt{|H|}} \left( {K\over 2}
\right)^{g-{1\over 2}}
\!\!\!\!
\sign{P} e^{i{\pi\over 2}\sign{H\over P}}
\ebst
\nonumber\\
&&\qquad\times
\ecass
\nonumber\\
&&\qquad\qquad\times
\sum_{\xl=0}^{\infty}{1\over \xl!} \left({1\over 8\pi iK} {P\over H}
\right)^{\xl}
\partial_\phi^{(2\xl)}\left. \frac
{\pjn \sin\left({2\pi\over p_j}(r_j\xn_j - \phi)\right) }
{\denphi}\right|_{\phi = {\bst\over 2K}},
\label{3.1.38}\\
\zirr & = &
\frac{(-1)^{n(l+1)} i\pi}{\pjn \Szpm{\xnp_j\over p_j}}
\left({K\over 2}\right)^{g-1}
\frac{\sign{P}}{\sqrt{|P|}} \eis
\nonumber\\
&&\times
\xep
\nonumber\\
&&\times
\ecass
\nonumber\\
&&\times
\left\{
{i\over 2} (-2)^n \Res_{\b=0} \left\{
\frac{\zeb}{\den} \cot(\pi\b)
\right.
\right.
\nonumber\\
&&\times
\pjn \left[
i \sin \left[ 2\pi \rpnsl \right]
\cos \left( {\pi\over K} {\b\p \over p_j}\right)
\sin \left( 2\pi\b\p {\xnp_j\over p_j} \right)
\right.
\nonumber\\
&&\left.\left.\qquad
+ \cos \left[ 2\pi \rpnsl \right]
\sin \left( {\pi\over K} {\b\p \over p_j}\right)
\cos \left( 2\pi\b\p {\xnp_j\over p_j} \right)
\right]\right\}
\nonumber\\
&&  - \smpj \sms
\nonumber\\
&&\times
\Res_{\b=0} \frac {\zebl} {\den}
\nonumber\\
&&\left.
\qquad\times
\pjn \mu\p_j \snrp
\right\}.
\label{3.1.39}\\
\zirsp & = & \zirr + \frac{(-1)^{n(l+1)} K^g}
{2^{n+3g-3} \pi^{n+2g-2}}
\frac{\sign{P}}{\sqrt{|P|}}
\frac{\eis}{\pjn\Szpm{\xnp_j\over p_j}}
\nonumber\\
&&\times
\xep
\nonumber\\
&&\times
\ecass
\nonumber\\
&&\times
\sum_{\xl=0\atop \xl-n\in 2 \ZZ}^{\infty}
\left( {1\over 2\pi i K}{P\over H} \right)^{\xl-n-2g+3\over 2}
\frac{\G{\xl-n-2g+3\over 2}}{\xl!}
\nonumber\\
&&\times
\left.
\partial_\phi^{\xl}
\left\{
\left( {2\pi\phi \over \sin(2\pi\phi)} \right)^{n+2g-2}
\!\!\!\!\!
\sum_{\mu\p_j = \pm 1 \atop \smunp \in \ZZ}
\pjn
\sin\left[ 2\pi\left({r_j\mu\p_j\xnp_j - \phi\over p_j}
-{1\over 2}s_j l\right)\right]
\right\}
\right|_{\phi=0}
\label{3.1.40}
\qqq
\end{proposition}

A condition $\sjn {\pm\xnp_j\over p_j}\not\in \ZZ$ in the second sum
of eq.~(\ref{3.1.37}) means that for any choice of signs $\pm$ in
front of the numbers $\xnp_j$ the sum is never integer. The condition
$\exists\mu\p_j=\pm 1:\;\smunp\in\ZZ$ means on the contrary that
there exists a choice of signs such that the sum is integer.

\nsection{Flat Connections and Asymptotic Contributions}
\label{s4}
\subsection{Connected Components of Moduli Space}
\label{ss4.1}
Our goal is to relate the terms $\zred$, $\zirr$ and $\zirsp$ of
the asymptotic formula~(\ref{3.1.37}) to connected components of
the moduli space $\cM(\Xgn)$ of flat connections of the Seifert
manifold $\Xgn$ in accordance with the quantum field theory
prediction~(\ref{1.10}). In this subsection we describe the connected
components of $\cM(\Xgn)$.

A flat connection $A_\mu$ on a manifold induces a homomorphism
$\hola$ of the fundamental group $\pi_1$ into the gauge group which
in our case is $SU(2)$. This homomorphism maps an element $x\in\pi_1$
into a parallel transport along $x$:
\qq
\hola(x) = \PexpA{x} \in SU(2).
\label{2.2.1}
\qqq
Two flat connections $A_\mu$ and $A_\mu\p$ are gauge equivalent iff
there exists an element $h$ of the gauge group which conjugates one
homomorphism into another:
\qq
\holap = h^{-1} \hola h.
\label{2.2.2}
\qqq
Therefore $\cM$ is also a moduli space of homomorphisms
$\pi_1\rightarrow SU(2)$ up to a global conjugation.

The Seifert manifold $\Xgn$ is constructed by the surgeries
$U^{(p_i,q_i)}$ on the loops $P_j\times S^1$ of the manifold
$\Sigma_g\times S^1$ as it was described in Section~\ref{s2}. The
fundamental group of $\Xgn$ is generated by the following elements:
the loop $b$ along $S^1$, the loops $a_1,\ldots,a_n$ around $n$
punctures $P_j$ on $\Sigma_g$ and the standard generators
$c_1,d_1,\ldots, c_g,d_g$ of $\pi_1(\Sigma_g)$. These elements
satisfy relations
\qq
&
a_j^{p_j} b^{q_j} = 1,\qquad 1\leq j\leq n,
&
\label{2.2.3}\\
&
a_1\cdot\ldots\cdot a_n = c_1 d_1 c_1^{-1} d_1^{-1}\cdot\ldots\cdot
c_g d_g c_g^{-1} d_g^{-1}
&
\label{2.2.4}
\qqq
\
and the requirement that $b$ commutes with all other elements of
$\pi_1$.

There is another set of important elements in $\pi_1$. These
elements represent the middle cycles of the solid tori (\ie their
parallels) which we glued in during the surgeries:
\qq
f_j = a_j^{r_j} b^{s_j}, \qquad 1\leq j \leq n.
\label{2.2.04}
\qqq

Consider a homomorphism $\hola: \pi_1\rightarrow SU(2)$. We introduce
a function $\phi: \pi_1\rightarrow [0,{1\over 2}]$ such that for
$x\in \pi_1$ both $\hola(x)$ and $\exp[2\pi i\sigma_3 \phi(x)]$
belong to the same conjugation class of $SU(2)$. Since $b$ commutes
with $a_j$, eqs.~(\ref{2.2.3}) and~(\ref{2.2.04}) imply that for some
numbers $\xnt,\xnt\p\in\ZZ$
\qq
\phi(a_j)
& = &
\left|{\xnt_j + q_j \phi(b) \over p_j} \right|,
\label{2.2.5}\\
\phi(f_j)
& = &
\left| \frac{\phi(b) - r_j\xnt_j}{p_j} + \xnt\p_j
\right|.
\label{2.2.05}
\qqq

The remaining analysis depends on the value of $\phi(b)$. If
$\phi(b) \neq 0,{1\over 2}$, then $\hola(b)$ does not belong to the
center of $SU(2)$. Therefore since $b$ belongs to the center of
$\pi_1(\Xgn)$, all the holonomies should belong to the same $U(1)$
subgroup of $SU(2)$, in particular,
\qq
\hola(b) = \exp [2\pi i\sigma_3 \phi(b)],\qquad
\hola(a_j) = \exp \left(
2\pi i \sigma_3
{\xnt_j + q_j \phi(b) \over p_j} \right).
\label{2.2.6}
\qqq
This means that the connection is reducible: the isotropy group
$H_c$, which commutes with the holonomies, is equal to $U(1)$. Also
since all the holonomies now commute, the \rhs of eq.~(\ref{2.2.4})
is trivial.  Therefore for some $\xmt\in \ZZ$
\qq
\xmt + \sjn
{\xnt_j + q_j \phi(b) \over p_j} = 0.
\label{2.2.7}
\qqq
Substituting here eq.~(\ref{2.2.5}) we find that
\qq
\phi(b) = {P\over H} \left( \xmt - \sjn {\xnt_j\over p_j} \right).
\label{2.2.8}
\qqq
As for the phases $\phi(c_j),\phi(d_j)$, $1\leq j \leq g$, they are
totally unrestricted. The only condition on $\hola(c_j)$ and
$\hola(d_j)$ is that they belong to the same subgroup $U(1)\subset
SU(2)$ as all other holonomies.
\begin{proposition}
\label{p2.2.1}
The connected components of reducible flat connections with $\phi(b)
\neq 0,{1\over 2}$ are $\mred$. Their holonomies are described by
eqs.~(\ref{2.2.6}),~(\ref{2.2.8}) and~(\ref{2.2.5}). The choice of
numbers $\xnt_1,\ldots,\xnt_n,\xmt$ is limited by a condition
\qq
0\leq \phi(a_1),\ldots,\phi(a_n),\phi(b)\leq {1\over 2}.
\label{2.2.9}
\qqq
\end{proposition}

If $\phi(b)=0,{1\over 2}$ then $\hola(b)$ belongs to the center of
$SU(2)$ and the connection can be irreducible. Eqs.~(\ref{2.2.3})
restrict the possible conjugation classes of the holonomies
$\hola(a_j)$. Since this time $\hola(b)$ is invariant under the
reflection
$e^{2\pi i\phi\sigma_3} \rightarrow e^{-2\pi i\phi\sigma_3} $,
we find that
\qq
\phi(a_j) = {\xnt_j + q_j\phi(b)\over p_j}.
\label{2.2.09}
\qqq
If $g=0$, then eq.~(\ref{2.2.4}) degenerates into
\qq
a_1\cdot\ldots\cdot a_n = 1.
\label{2.2.10}
\qqq
This condition imposes a quantum group version of the polygon (\eg,
triangle for $n=3$) inequalities on the phases $\phi(a_j)$. If
however $g\geq 1$, then since the commutants $h_1 h_2 h_1^{-1}
h_2^{-1}$, $h_{1,2}\in SU(2)$ cover the whole group $SU(2)$,
eq.~(\ref{2.2.4}) does not restrict the phases $\phi(a_j)$.
\begin{proposition}
\label{p2.2.2}
The connected components of irreducible flat connections are $\mirr$.
The conjugation classes of some their holonomies are determined by
eq.~(\ref{2.2.09}) with $\phi(b) = {\tilde{l}\over2},\;\;
\tilde{l}=0,1$.  The choice of the numbers $\xnt_j$ is limited by the
condition~(\ref{2.2.9}).

If there exist the numbers $\tilde{\mu}_j=\pm 1$ such that
$\sjn \tilde{\mu}_j \phi(a_j) \in \ZZ$, (cf. eq.~(\ref{2.2.7}))
then
some of the connections of the connected component $\mirr$ are
reducible and we denote it as $\mirsp$.
\end{proposition}

\subsection{Identification of Asymptotic Contributions}
\label{ss4.2.1}
We are going to identify the contributions that the connected
components of the moduli space $\cM(\Xgn)$ make to Witten's invariant
$\zx$.
\begin{proposition}
\label{p4.2.1}
The contribution to Witten's invariant $\zx$ of a reducible component
$\mred$ is $\zred$ of eq.~(\ref{3.1.38}) such that
\qq
\xn_j = \xnt_j (\mod p_j),\qquad
\xm = \xmt + \sjn{\xn_j - \xnt_j\over p_j}.
\label{4.2.1}
\qqq
The contribution of an irreducible component $\mirr$ is $\zirr$ such
that
\qq
\xnp_j = \xnt_j + {1\over 2}q_j \tilde{l}, \qquad
l = \tilde{l}.
\label{4.2.2}
\qqq
The contribution of a special irreducible component $\mirsp$ (which
also contains some reducible connections) is $\zirsp$ whose indices
are given by  eq.~(\ref{4.2.2}).
\end{proposition}

One possible way of verifying these claims is to use
eqs.~(\ref{1.11}) and~(\ref{1.14}). One has to compare the already
known Chern-Simons actions of flat connections to the leading
exponentials of eqs.~(\ref{3.1.38})-(\ref{3.1.40}). One-loop
corrections can also be compared if at least some of the parameters
in the \rhs of eq.~(\ref{1.14}) can be independently calculated. We
carried out this program for 3-fibered Seifert manifolds
$X_0\left({p_1\over q_1},{p_2\over q_2},{p_3\over q_3}\right)$
in~\cite{Ro1} by using the 1-loop calculations of~\cite{FrGo}.

A more direct way of identifying the asymptotic contributions is to
``measure'' (or, in the language of quantum theory, ``observe'')
directly the holonomies of flat connections along some elements of
the fundamental group of the manifold. Suppose that we know that for
an $x\in\pi_1$ the conjugation class of $\hola(x)$ is the
same for all connections of a connected component $\cM_c$. Let us
introduce a knot (that is, a Wilson line) along $x$ carrying a
$\gamma$-dimensional representation of $SU(2)$. In other words, we
multiply the integrand of eq.~(\ref{1.2}) by an extra factor
$\Tr_\gamma\PexpA{x}$. According to eq.~(\ref{1.14}), at the 1-loop
level in $1/K$ expansion the contribution of $\cM_c$ will be
multiplied by
\qq
\Tr_\gamma\PexpA{x} = \Tr_\gamma \exp[2\pi i\sigma_3\phi(x)]
\equiv \frac{\sin[2\pi\gamma\phi(x)]}{\sin[2\pi\phi(x)]}.
\label{4.2.3}
\qqq
Therefore the knot is an observable which measures the conjugation
class of the holonomy.

We introduce the following link into the Seifert manifold $\Xgn$:
a
line along $b$ with $\gamma$-dimensional representation and $n$ lines
along $a_j$ with $\gamma_j$ dimensional representations. The new
Witten's invariant $\zgx$ can be easily calculated with the help of
the lemma whose simple proof can be traced back to~\cite{Wi1}:
\begin{lemma}
\label{l4.2.1}
Let $\cK$ be a knot in a manifold $M$ and let $\cK_m$ be the meridian
of $\cK$. If $\cK$ carries an $\a$-dimensional representation and
$\cK_m$ carries a $\gamma$-dimensional representation, then
\qq
Z_{\a,\gamma}(M,\cK,\cK_m;k) =
{\tS^{-1}_{\gamma\a}\over \tS^{-1}_{1\a}}
Z_{\a}(M,\cK;k)
\equiv
\frac{\sin\left({\pi\over K}\a\gamma\right)}
{\sin\left({\pi\over K}\a\right)}
Z_{\a}(M,\cK;k).
\label{4.2.4}
\qqq
\end{lemma}
As a result,
\qq
\zgx & = & e^{i\ffr} \sum_{\a_j=1}^{K-1}
N^g_{\a_1,\ldots,\a_n,\gamma} \pjn
\frac
{\sin\left({\pi\over K}\a_j\gamma_j\right)}
{\sin\left({\pi\over K}\a_j\right)}
\tU^{(p_j,q_j)}_{\a_j 1}
\label{4.2.5}\\
& = &
e^{i\ffr} \sum_{\b=1}^{K-1}
\frac
{\sin\left({\pi\over K}\b\gamma\right)}
{\sin\left({\pi\over K}\b\right)}
\frac
{\pjn\sum_{\a_j=1}^{K-1}
\frac
{\sin\left({\pi\over K}\a_j\gamma_j\right)}
{\sin\left({\pi\over K}\a_j\right)}
\tS_{\b\a_j}\tU^{(p_j,q_j)}_{\a_j 1}}
{\tS^{n+2g-2}_{\b 1}}.
\nonumber
\qqq

Instead of going through the detailed asymptotic calculation of the
sums of this equation along the lines of the previous section (which
is possible but tedious) we will present a simple argument which will
show how the extra factors
\qq
\frac
{\sin\left({\pi\over K}\b\gamma\right)}
{\sin\left({\pi\over K}\b\right)}
\label{4.2.5f1}
\qqq
and
\qq
\frac
{\sin\left({\pi\over K}\a_j\gamma_j\right)}
{\sin\left({\pi\over K}\a_j\right)}
\label{4.2.5f2}
\qqq
affect the asymptotic formulas~(\ref{3.1.38}) and~(\ref{3.1.40}).
Note that all the terms in eq.~(\ref{3.1.37}) came as local
contributions of some special points $\b^\ast$: $\zred$ came
from the stationary phase points $\b^\ast = \bst$, $\zirr$ came from
residues at $\b^{\ast} = Kl$ and $\zirsp$ came from both stationary
phase and residue at $\b^{\ast} = Kl$. Therefore to the leading order
in $K$ the effect of the factor~(\ref{4.2.5f1})
is to multiply these contributions by
$
\frac
{\sin\left({\pi\over K}\b^{\ast}\gamma\right)}
{\sin\left({\pi\over K}\b^{\ast}\right)}
$.
Comparing this factor with the \rhs of eq.~(\ref{4.2.3}) we conclude
that
\qq
\phi(b) = {\b^\ast\over 2K}.
\label{4.2.6}
\qqq
This means that for $\zred$
\qq
\phi(b) = {P\over H} \left( \xm - \sjn {\xn_j\over p_j} \right),
\label{4.2.7}
\qqq
while for $\zirr$ and $\zirsp$
\qq
\phi(b) = {l \over 2}
\label{4.2.8}
\qqq
in full agreement with the Proposition~\ref{p4.2.1}.

To find the effect of the factor~(\ref{4.2.5f2})
consider the calculation of the sum
\qq
\sum_{\a_j = 1}^{K-1} \tS_{\b\a_j} \tU^{(p_j,q_j)}_{\a_j 1},
\qqq
which produces the factor $\tU^{(-q_j,p_j)}$ of eq.~(\ref{2.1.7}). The
relevant part of this sum is
\qq
\sum_{\a_j=1}^{K-1}\exp \left[ {i\pi\over 2K} \left(
{p_j\over q_j}\a_j^2 - 2\a_j \left( {2K\xn_U + \mu_U\over q_j}
+ \b\mu_S\right)\right)\right],
\label{4.2.9}
\qqq
here $\xn_U$ and $\mu_U$ are $\xn$ and $\mu$ coming from
eq.~(\ref{1.7}) while $\mu_S$ comes from the formula
\qq
\sin\left( {\pi\over K} \b\a_j\right) = {i\over 2}
\sum_{\mu_S = \pm 1} \mu_S \exp \left(-i{\pi\over K}
\mu_S \b\a_j\right).
\qqq
The sum~(\ref{4.2.9}) can be calculated along the lines of the
previous section. It will turn into a purely gaussian integral over
$\a_j$. The stationary phase point which dominates this integral is
\qq
\ajst = \frac{2K\xn_U + \mu_S q_j \b}{p_j}.
\label{4.2.10}
\qqq
On the other hand, comparing an integral over $\a_j$ of the summand
in eq.~(\ref{4.2.9}) with the exponentials of eq.~(\ref{2.1.8}) we
conclude that
\qq
\xn_j = \mu_S \xn_U,\qquad
\mu_j = \mu_S \mu_U,
\label{4.2.11}
\qqq
so that
\qq
\ajst = \mu_S\frac{2K\xn_j + q_j\b}{p_j}.
\label{4.2.12}
\qqq
Therefore to the leading order in $K$, the effect of the
factor~(\ref{4.2.5f2}) is to multiply the contributions by
\qq
\frac
{\sin\left({\pi\over K}\ajst\gamma_j\right)}
{\sin\left({\pi\over K}\ajst\right)},
\qqq
with $\ajst$ coming from eq.~(\ref{4.2.12}) in which
we should substitute $\b=\b^\ast$. Then eq.~(\ref{4.2.3}) tells us
that
\qq
\phi(a_j) = \left| \frac{\xn_j + q_j{\b^\ast\over 2K}}{p_j}
\right|,
\label{4.2.13}
\qqq
which is again in full agreement with the Proposition~\ref{p4.2.1}.

Finally as a result of our identifications we can recognize the
presence of the factors
$
%
\pjn \sin [2\pi\phi(f_j)]
%
$
in all the formulas~(\ref{3.1.38})-(\ref{3.1.40}), \eg the factors
$
\pjn\sin\left({2\pi\over p_j}(r_j\xn_j - \phi)\right)
$
in eq.~(\ref{3.1.38}) and
$
\pjn\sin\rpnsl
$
in eq.~(\ref{3.1.39}).

\nsection{Intersection Numbers on Moduli Space}
\label{s5}
Consider again the asymptotic formulas~(\ref{3.1.37})-(\ref{3.1.40}).
Whereas the contributions of reducible connections $\zred$ are
presented as infinite asymptotic series in $1/K$, it turns out that
the contributions of irreducible connections are in fact finite
polynomials in $1/K$. This follows easily from the residue
formula~(\ref{3.1.39}). The situation seems similar to that of the
Yang-Mills partition fuction calculation of ~\cite{Wi3} and the
calculation of Verlinde numbers in ~\cite{Sz1},~\cite{BeSz}. In all
these cases the moduli spaces contributing the polynomials to the
partition functions are isomorphic. In particular, it is easy to see
that
\qq
\mirr(\Xgn) = \mts, \qquad
\theta_j = {\xnt_j + {1\over 2}q_j \tilde{l} \over p_j}
\equiv {\xnp_j\over p_j},
\label{5.1.1.}
\qqq
here $\mts$ is a moduli space of $SU(2)$ flat connections of a
$g$-handle surface with $n$ punctures $P_j$ and holonomies around
them fixed by eq.~(\ref{2.1.11}). Both the Yang-Mills partition
function~(\ref{2.1.12}) and Verlinde number~(\ref{2.1.10}) were
expressed in terms of the intersection numbers on $\mts$. We will
derive a similar expression for $\zirr$ by comparing the asymptotic
formulas for these three objects and using the localization formulas
of~\cite{Wi3},~\cite{JeKi} and~\cite{Sz2}.

We start by presenting the residue formulas for the partition
functions~(\ref{2.1.10}) and~(\ref{2.1.13}).
\begin{proposition}
\label{p5.1.1}
A number of conformal blocks for the $SU(2)$ WZW model on $\Sigma_g$
with $n$ insertions of the primary fields $\cO_{\a_j}$ is equal to
\qq
\nga & = & -4\pi \left({K\over 2}\right)^g \left[
\Res_{\phi=0}\frac
{\pjn \sin(2 \pi \a_j \phi)} {\denphi} \cot(2\pi K\phi)
\right.
\label{5.1.2}\\
&&\qquad
-\fri \smj \pmj \sign{\smua}
\sum_{0\leq \xxm <{1\over 2K}|\smua|}
{1\over \spm{\xxm}}
\nonumber\\
&&\qquad\qquad\times
\left.
\Res_{\phi=0} \frac
{\exp \left[ 2\pi i\phi \sign{\smua} \left( 2K\xxm -
\left|\smua\right| \right)\right]}
{\denphi} \right],
\nonumber
\qqq
if $n+2g-2>0$ and $n+\sjn \a_j$ is even. If $n+\sjn\a_j$ is odd then
$\nga=0$.
\end{proposition}


\begin{proposition}
\label{p5.1.2}
A partition function of the 2d Yang-Mills theory on a $g$-handled
surface $\Sigma_g$ with $n$ punctures $P_j$, the holonomies around
which are fixed by eqs.~(\ref{2.1.11}), has the following asymptotic
representation in the limit of small gauge coupling constant $a$: if
$\sjn\pm\theta_j\not\in\ZZ$ then
\qq
\zt & = & \ztirr + \smj \sum_{\xxm\in \ZZ \atop \xxm -\smut >0}
\ztred,
\label{5.1.3}\\
\ztirr & = & - {1\over 2^g \pi^{n+2g-3}}
\left[
\Res_{\phi=0}\frac
{e^{-a\phi^2}}{\phi^{n+2g-2}} \cot(\pi\phi)
\pjn \sin(2\pi\theta_j\phi)
\right.
\label{5.1.4}\\
&&\qquad
-\fri
\smj \pmj \sum_{0\leq\xxm < \left|\smut\right|}
{\sign{\smut}\over \spm{\xxm}}
\nonumber\\
&&\qquad\times\left.
\Res_{\phi=0}\frac
{\exp \left[ -a\phi^2 + 2\pi i \phi \sign{\smut}
\left(\xxm - \left|\smut\right|\right)\right]}
{\phi^{n+2g-2}}\right],
\nonumber\\
\ztred & = &
\pmj \exp \left[ -{\pi^2\over a}\left(\xxm - \smut\right)^2\right]
\sum_{\xxl=0}^{\infty} \frac
{(-1)^{g-1+\xxl} a^{n+2g+\xxl - {5\over 2}}}
{2^{n+2g+2\xxl} \pi^{2n + 4g + 2\xxl - {9\over 2}} \xxl!}
\label{5.1.5}\\
&&\qquad\times
\frac{(n+2g+2\xxl-3)!}{(n+2g-3)!}
{1\over \left(\xxm - \smut\right)^{n+2g+\xxl -2}}.
\nonumber
\qqq

If $n$ is even and
$\exists\mu_j=\pm 1$ such that $\smut\in\ZZ$, then $\ztirr$ in
eq.~(\ref{5.1.3}) should be substituted by $\ztirsp$:
\qq
\ztirsp & = & \ztirr + \frac
{i^n a^{2n+2g-3\over 2}} {2^{n+g}\pi^{n+2g-2}}
\G{3-2g-n\over 2}
\sum_{\mu_j = \pm 1 \atop \smut\in\ZZ}
\pmj
\label{5.1.6}\\
& \equiv &  \ztirr -
(-1)^{g + {n\over 2}}
\frac{i^n 2^{g-2}}{\pi^{n + 2g - {5\over 2}}}
\frac{\left({n+2g-2\over 2}\right)!} {(n + 2g - 2)!}
a^{n+g-{3\over 2}}
\sum_{\mu_j = \pm 1 \atop \smut\in\ZZ}
\pmj
\nonumber
\qqq
\end{proposition}
The contributions $\ztred$ come from constant curvature $U(1)$
connections, the contribution $\ztirr$ comes from irreducible flat
connections.

Eq.~(\ref{5.1.2}) was derived (for the case of $n=0$) in the
papers~\cite{Sz1},~\cite{BeSz}. E.~Witten derived eqs.~(\ref{5.1.5})
and eq.~(\ref{5.1.6}) in~\cite{Wi3}.

According to~\cite{Wi3},
\qq
2\ztirr = \int_{\mts} \exp\left(\omega +
4a
\Theta\right),
\label{5.1.8}
\qqq
here $\Theta$ is a 4-form defined in~\cite{Wi3}
and $\omega$ is a
symplectic form on $\mts$ normalized in the following way: if $a_\mu$
and $b_\mu$ are two $su(2)$ valued 1-forms representing the tangent
vectors at a point on $\mts$ then
\qq
\omega(a_\mu,b_\mu) = {1\over
4\pi^2} \Tr\int_{\Sigma_g} a_\mu\wedge b_\mu.
\label{5.1.9}
\qqq

The moduli space $\mts$ is a bundle over a moduli space $\ms$ of flat
connections on $\Sigma_g$ without punctures\footnote{
I am thankful to L.~Jeffrey and A.~Szenes for explaining to me the
properties of this bundle and its symplectic structure.}
(see \cite{Do},
let us forget for a
moment that $\ms$ has a singularity, we also assume that $\theta_j$
are small and $\sjn\pm\theta_j\not\in\ZZ$). The symplectic form
$\omega$ is a sum of forms
\qq
\omega = \omega_0 + 2\sjn\theta_j\omega_j,
\label{5.1.11}
\qqq
here $\omega_0$ is a pull-back
of the symplectic form on $\ms$ while
$\omega_j$ are closed 2-forms normalized so that
\qq
\int_{S^2_i}\omega_j = \delta_{ij},
\label{5.1.12}
\qqq
$S_i^2$ $(q\leq i\leq n)$ are the 2-dimensional spheres which make up
the fibers of the bundle $\mts\rightarrow\ms$.

The Verlinde number~(\ref{5.1.2}) is a dimension of the Chern-Simons
Hilbert space for $\Sigma_g$ with $n$ insertions of primary fields
$\cO_{\a_j}$. In other words, it is a number of holomorphic sections
of a certain line bundle over $\mts$ with
\qq
\theta_j={\a_j-1\over 2k}.
\label{5.1.13}
\qqq
Therefore it is given by the Riemann-Roch formula
\qq
\nga = \int_{\mts} e^{k\omega}\Td(\mts)
\label{5.1.14}
\qqq
(see, \eg~\cite{Wi2},~\cite{Sz2} and references therein). Note that a
natural symplectic
form coming from eqs.~(\ref{1.1}) and~(\ref{1.2}) is
\qq
\omega\p = 4\pi^2 \omega.
\label{5.1.10}
\qqq
Therefore the semiclassical formula for the dimension of the Hilbert
space should contain the exponent $\exp\left({\omega\p\over
2\pi\hbar}\right) = \exp\left({k\omega\p\over 4\pi^2}\right)$ in full
agreement with eq.~(\ref{5.1.14}).

The Todd class $\Td(\mts)$ can be expressed as
\qq
\Td(\mts) = \exp\left(2\omega_0 + \sjn \omega_j\right)
\ahat(\mts)
\label{5.1.15}
\qqq
(see, \eg~\cite{Sz2} and references therein). Upon substituting this
expression in eq.~(\ref{5.1.14}) we get
\qq
\nga = \int_{\mts} \exp \left( K\omega_0 + \sjn \a_j\omega_j \right)
\ahat(\mts).
\label{5.1.16}
\qqq
The pairs of equations~(\ref{5.1.2}),~(\ref{5.1.16})
and~(\ref{5.1.4}),~(\ref{5.1.8}) are particular cases of the
following conjecture which can be deduced from the calculations
of~\cite{Wi3}, the main theorem of~\cite{JeKi} and the calculations
and conjecture of~\cite{Sz2}:
\begin{conjecture}
\label{c5.1.1}
For the numbers $\theta_j,1\leq j\leq n$ such that $\sjn\pm
\theta_j\not\in \ZZ$ let $\mts$ be the moduli space of flat $SU(2)$
connections on $\Sigma_g$ with $n$ punctures and
holonomies~(\ref{2.1.11}) around them. Then for the two (not
necessarily integer) numbers $K,a$
\qq
\lefteqn{
\int_{\mts}\exp\left[K\left(\omega_0 + 2\sjn\theta_j\omega_j\right)
+
4a
\Theta\right] \ahat(\mts)
}
\label{5.1.17}\\
&=&-2\pi\left({K\over 2}\right)^g
\left[
\Res_{\phi=0}
\frac
{\exp(-a\phi^2)}{\denp}\cot(\pi K\phi)\pjn\sin(2\pi K\theta_j \phi)
\right.
\nonumber\\
&&
-\fri \smj \pmj \sign{\smut} \sum_{0\leq \xxm < \left|\smut\right|}
{1\over \spm{\xxm}}
\nonumber\\
&&\left.
\qquad\times
\Res_{\phi=0}
\frac
{\exp\left[ -a\phi^2 + 2\pi iK\phi\sign{\smut}
\left(\xxm-\left|\smut\right|\right)\right]}
{\denp}\right].
\nonumber
\qqq
\end{conjecture}

Suppose that we change the phases $\theta_j$ by small amounts
$\Delta\theta_j$ such that for any $t\in [0,1]$ $\sjn \pm(\theta_j +
t\Delta\theta_j) \not \in \ZZ$. The topological class of the
manifold $\mts$ does not change. As a result,
\qq
\lefteqn{
\int_{\mts}\exp\left[K\left(\omega_0 +
2\sjn(\theta_j + \Delta\theta_j)   \omega_j\right)
+
4a
\Theta\right] \ahat(\mts)
}
\label{5.1.18}\\
&=&-2\pi\left({K\over 2}\right)^g
\left[
\Res_{\phi=0}
\frac
{\exp(-a\phi^2)}{\denp}\cot(\pi K\phi)
\pjn\sin(2\pi K(\theta_j + \Delta\theta_j) \phi)
\right.
\nonumber\\
&&
-\fri \smj \pmj \sign{\smut} \sum_{0\leq \xxm < \left|\smut\right|}
{1\over \spm{\xxm}}
\nonumber\\
&&\qquad\times\left.
\Res_{\phi=0}
\frac
{\exp\left[ -a\phi^2 + 2\pi iK\phi\sign{\smut}
\left(\xxm-\left|\sjn\mu_j(\theta_j+\Delta\theta_j)
\right|\right)\right]}
{\denp}\right].
\nonumber
\qqq

It is easy to put the \rhs of eq.~(\ref{3.1.39}) in a form similar to
the \rhs of eq.~(\ref{5.1.18})
for the case when $\sjn\pm{\xnp_j\over p_j}\not\in\ZZ$:
\qq
\zirr & = &
-
\frac{(-1)^{nl} \pi}{\pjn \Szpm{\xnp_j\over p_j}}
\left({K\over 2}\right)^{g}
\frac{\sign{P}}{\sqrt{|P|}} \eis
\label{5.1.19}\\
&&\times
\xep
\nonumber\\
&&\times
\ecass
\nonumber\\
&&\times
\smj \pmj
\yep
%
%
\nonumber\\
&&\times
\left\{
\Res_{\phi=0}
\frac{\zeph}
{\denp} \cot(\pi K\phi)
\pjn\sin\left[2\pi K\phi\fnpp\right]
\right.
\nonumber\\
&&  - \fri \smpj \pmup \smsold
\nonumber\\
\lefteqn{
\!\!\!\!\!\!\!\!\!\!\!\!\!\!\!
\times\left.
\Res_{\phi=0} \frac {\zephl} {\denp}
\right\}.
}
\nonumber
\qqq
Comparing this expression with the intersection number
formula~(\ref{5.1.18}) we come to the following conclusion:
\begin{proposition}
\label{p5.1.3}
The contribution of a connected component $\mirr$ of the moduli space
of irreducible flat connections to Witten's invariant $\zx$ can be
expressed in terms of the intersection numbers of the forms on this
component:
\qq
\zirr & = &
\frac{(-1)^{nl}}{2}
\frac{\eis}{\pjn \Szpm{\xnp_j\over p_j}}
\frac{\sign{P}}{\sqrt{|P|}}
\label{5.1.20}\\
&&\times
\xep
\nonumber\\
&&\times
\ecass
\nonumber\\
&&\times
\smj \pmj
\yep
\nonumber\\
&&\times
\int_{\mirr}
\exp \left[ K \left( \omega_0 + 2\sjn \frac{\xnp_j + {\mu_j\over 2K}}
{p_j} \omega_j +
2
\pi i{H\over P} \Theta \right)\right]
\ahat(\mirr),
\nonumber
\qqq
or, equivalently,
\qq
\zirr & = &
(-1)^{nl} 2^{n-1}
\frac{\eis}{\pjn \Szpm{\xnp_j\over p_j}}
\frac{\sign{P}}{\sqrt{|P|}}
\label{5.1.21}\\
&&\times
\xep
\nonumber\\
&&\times
\ecass
\nonumber\\
&&\times
\int_{\mirr}
\exp \left[ K \left( \omega_0 + 2\sjn \frac{\xnp_j}{p_j}
\omega_j +
2
\pi i{H\over P} \Theta \right)\right]
\ahat(\mirr)
\nonumber\\
\lefteqn{\hspace*{-1in}
\times
\pjn\left\{
i\cosh\left({\omega_j\over p_j}\right) \sin\left[2\pi\rpnsl\right]
+ \sinh\left({\omega_j\over p_j}\right) \cos\left[2\pi\rpnsl\right]
\right\}
}
\nonumber
\qqq
The numbers $\xnp_j$ and $\xnt_j$ are related by eq.~(\ref{4.2.2}),
also $\mirr$ is isomorphic to $\cM_{\{{\xnp\over p}\}}(\Sigma_g)$.
\end{proposition}

The formula~(\ref{5.1.20}) looks very similar to eq.~(\ref{5.1.8})
and also to eq.~(\ref{5.1.14}) if we recall that
\qq
N^g_{\{\a\}} = Z_{\{\a\}}(\Sigma_g\times S^1,\cL;k),
\label{5.1.22}
\qqq
the $n$-component link $\cL$ consists of $n$ loops which go along
$S^1$ of $\Sigma_g\times S^1$. E.~Witten proved eq.~(\ref{5.1.8})
in~\cite{Wi3} by applying the equivariant localization arguments to
the path integral representation of the 2d Yang-Mills theory. It
seems likely that there should be a path integral localization proof
for eq.~(\ref{5.1.20}) as well. We came to eq.~(\ref{5.1.20}) through
the back door: by working out the large $k$ asymptotics of the
surgery formula and then cooking up an intersection number that would
match the contribution of an irreducible connection. A localization
argument would derive the \rhs of eq.~(\ref{5.1.20}) directly from
the path integral~(\ref{1.2}). Note, however, that even for the
seemingly simpler case of eq.~(\ref{5.1.14}) there is no localization
proof yet. M.~Blau and R.~Thompson~\cite{BlTh} could only use abelian
localization in order to establish Verlinde formula~(\ref{2.1.4}). At
present time in order to prove the formula~(\ref{5.1.14}) one has to
show that the path integral for
$Z_{\{\a\}}(\Sigma_g\times S^1,\cL;k)$
is equal to the number of sections of a certain holomorphic line
bundle and then use the Riemann-Roch theorem to calculate that
number.

\nsection{Conclusion}
An extensive use of path integral arguments puts the theory of
Witten's invariants somewhere between mathematics and physics. The
path integral calculations are tested in physics against the
data coming from experiments with elementary particles.
In a similar way we can say that the asymptotic expansion of the
surgery formula~(\ref{1.5}) provides us with experimental data about
Seifert manifolds. This data has to be compared with the asymptotic
expansion~(\ref{1.12}) of the path integral.

Being viewed in this way, the annoying complexity of the
formulas~(\ref{3.1.37})-(\ref{3.1.40}) should be encouraging. It means
that there is plenty of experimental data (\ie topological invariants
of 3d manifolds) hidden in them. As we already know, this data
includes Chern-Simons invariants, Reidemeister-Ray-Singer torsion and
spectral flows at the 1-loop level. The Casson-Walker invariant
appears as a 2-loop correction to the contribution of the trivial
connection to Witten's invariant of rational homology spheres (and
Seifert manifolds $X_{0,\{{p\over q}\}}$ among them, see
\eg~\cite{Ro1}).  The full trivial connection contribution in the
general case of $\Xgn$ was studied in~~\cite{Ro2} with the help of
eq.~(\ref{3.1.40}). We do not repeat this analysis here.

In this
paper we were mostly interested in the contributions of irreducible
flat connections which appear to be finite loop exact. We were able
to express these contributions as certain intersection numbers on the
moduli space of flat connections. However eq.~(\ref{5.1.20}) was
derived ``through the back door'', that is, by comparing the residue
expression~(\ref{3.1.39}) coming from the surgery
formula~(\ref{5.1.18}) with the residue formula~(\ref{5.1.18}) for
the intersection numbers. It would be much better to derive
eq.~(\ref{5.1.20}) directly by applying some sort of localization
arguments in the spirit of~~\cite{Wi3} to the Chern-Simons path
integral~(\ref{1.2}). However this still remains an unsolved problem.

\section*{Acknowledgements}

I am thankful to L.~Jeffrey, A.~Szenes, A.~Vaintrob and E.~Witten for
valuable discussions and advice.

This work was supported by the National Science Foundation
under Grant No. PHY-92 09978.

\nappendix{1}
\label{a1}
There is an alternative way of calculating the sum~(\ref{3.1.1})
which is similar to the one used in~\cite{Ro1}. This method is a
Fourier transform of the method used in the Section~\ref{s3}. It
involves gaussian integrals instead of residues and boundary
contributions instead of stationary phase contributions.

We start by expanding the denominator of eq.~(\ref{3.1.1}) in an
analog of geometric series:
\qq
{1\over \denb} = (2i)^{n+2g-2} e^{-{i\pi\over K}(n+2g-2)(\b - i\xi)}
\sum_{\gamma\in\ZZ\atop \gamma\geq 0} K_{n+2g-2}(\gamma) e^{-{2\pi
i\over K} \gamma(\b - i\xi)}.
\label{3.2.1}
\qqq
Here $K_n(m)$ is the $SU(2)$ Kostant's partition function:
\qq
K_n(m) = {m + n - 1\choose n-1} \equiv
\frac{(m+n-1)!}{(n-1)!m!} = 2\pi i\Res_{x=0}
\frac{e^{2\pi i mx}}{(1-e^{-2\pi ix})^n}.
\label{3.2.2}
\qqq
In other words, the polynomial $K_n(m)$ is equal to the number of
ways in which an integer number $m$ can be split into a sum of $n$
ordered nonnegative intergers.

The expression~(\ref{3.2.1}) can be put in a different form if we use
a ``shifted'' Kostant's polynomial
\qq
\tK_n(m) & = & K_n ( m - {n\over 2} )  =
{1\over (n-1)!} \prod_{ 0\leq j\leq {n\over 2} -1 \atop j\in\ZZ +
{n\over 2}} (m^2 - j^2)^{1\over \spm{j}}
\label{3.2.3}\\
& = &
\frac{\pi}{(2i)^{n-2}} \Res_{x=0} \frac {e^{2\pi i mx}}{\sin^n(\pi
x)}
\qquad \hspace*{2in}
{\rm for}\;\;m\geq 0
\nonumber\\
\tK_n(m) & = & 0 \qquad {\rm for}\;\;m < 0.
\nonumber
\qqq
Since $K_n(m)=0$ (as defined by
$K_n(m) = {1\over (n-1)!} \prod_{j=1}^{n-1} (m+j)$) if $m \in \ZZ$,
$1-n\leq m \leq -1$, we can shift the range of summation in
eq.~(\ref{3.2.1}) so that
\qq
{1\over \denb} = (2i)^{n+2g-2} \sum_{\gamma>0\atop \gamma\in
\ZZ+{n\over 2}} \tK_{n+2g-2}(\gamma) \exp\left(-{2\pi i\over K}\gamma
(\b - i\xi)\right).
\label{3.2.4}
\qqq

The Poisson resummation formula
\qq
\sum_{m\in \ZZ + {n\over 2}} \delta(\gamma - m) =
\sl e^{\pi iln}\sum_{\b\p\in\ZZ}
e^{2\pi i(l-2\b\p)\gamma}
\label{3.2.5}
\qqq
allows us to convert the sum in eq.~(\ref{3.2.4}) into an integral
over $\gamma$:
\qq
{1\over \denb} & = &
(2i)^{n+2g-2} \sl e^{\pi iln} \int_0^{\infty} d\gamma
\tK_{n+2g-2}(\gamma)
\label{3.2.6}\\
&&\qquad\times
\exp\left( - {2\pi i\over K}
(\b + 2K\b\p - Kl - i\xi)\right).
\nonumber
\qqq
We substitute this expression into eq.~(\ref{3.1.1}). Since the
summand of eq.~(\ref{3.1.1}) is invariant under the shift
$\b\rightarrow \b+ 2K$, we can combine the sums
$\sum_{\b=-K+1}^{K}\sum_{\b\p\in\ZZ}$
into one sum $\sum_{\b\in\ZZ}$,
which we transform into an integral with the help of the Poisson
formula~(\ref{3.1.5}):
\qq
\zsx & = &
\frac{(2i)^{n+2g-2}}{2} \sl e^{i\pi ln} \snp
\smj \pmj \xe
\nonumber\\
&&\times
\lim_{\xi\rightarrow 0^+}
\sum_{\xm\in\ZZ}
\int_{0}^{\infty} d\gamma \tK_{n+2g-2}(\gamma)
\exp\left[2\pi i\gamma\left(l+{i\xi\over K}\right)\right]
\label{3.2.7}\\
&&\qquad\times
\int_{-\infty}^{+\infty} d\b
\exp \left[-{2\pi i\over K} \left({1\over 4}{H\over P}\b^2 +
\b\left(\gamma - K\xm + \sjn \frac{K\xn_j +{\mu_j\over 2}}{p_j}
\right)\right)\right].
\nonumber
\qqq
The integral over $\b$ is purely gaussian and straightforward to
calculate. We go from $\xn_j$ to $\xnp_j$ according to
eq.~(\ref{3.1.27}) and transform a sum $\snp$ into
\qq
\snpjh \smpj
\label{3.2.8}
\qqq
by substituting $\mu\p_j\xnp_j$ for $\xnp_j$. Since
$\tK_{n+2g-2}(\gamma) = 0$ for $\gamma<0$ we can extend the
integration range to all $\gamma$. We also substitute
\qq
\gamma + K \xm - \sjnmu
\label{3.2.9}
\qqq
for $\gamma$. After all these transformations we end up with the
following expression:
\qq
\zsx & = &
(2i)^{n+2g-2} \left( {K\over 2} \left|{P\over H}\right|
\right)^{1\over 2} e^{-{i\pi \over 4}\sign{H\over P}}
\sl e^{i\pi ln} \snpjh
\nonumber\\
&&\times
\xep
\label{3.2.10}\\
&&\times
\smj \pmj \yep
\nonumber\\
&&\qquad\times
\int_{-\infty}^{+\infty} \tKt \exp \left( {2\pi i\over K}{P\over H}
\gamma^2 \right),
\nonumber
\qqq
here
\qq
\tKt & = & \lim_{\xi\rightarrow 0^+}
\sum_{\xm\in\ZZ} \smpj \pmup
\label{3.2.11}\\
&&\times
\tK_{n+2g-2}\left(\gamma + K\xm - \sjnmu \right)
\nonumber\\
&&\qquad\times
\exp \left[-{2\pi\over K}\xi\left(\gamma + K\xm - \sjnmu \right)
\right]
\nonumber
\qqq
The function $\tKt$ is locally polynomial in $\gamma$ but it (or its
derivatives) has a break at the points
\qq
\gbr = - K\xm + \sjnmu,\qquad \xm \in \ZZ,
\label{3.2.12}
\qqq
because the shifted Kostant's partition function
$\tK_{n+2g-2}(\gamma)$ (or its derivatives) has a break at
$\gamma=0$.

The sum $\sum_{\xm\in\ZZ}$ in eq.~(\ref{3.2.11}) can be limited to
\qq
\xm \geq -{\gamma\over K} + \sjnmuk,
\label{3.2.13}
\qqq
because $\tK_{n+2g-2}(\gamma)=0$ if $\gamma<0$. The remaining
semi-infinite sum over $\xm$ is regularized by the factor
$e^{-2\pi\xm\xi}$ which is present in eq.~(\ref{3.2.11}). Actually, if
$g=0$, then the alternating sum over $\mu\p_j$ is similar to the ones
which express the weight multiplicities of tensor products through
Kostant's partition functions. Therefore the sums
\qq
\smpj \pmup \tK_{n-2} \left(\gamma + K\xm - \sjnmu \right)
\label{3.2.14}
\qqq
are equal to zero if $\gamma + K\xm$ is big enough. As a result, only
a finite number of terms contribute to the sum over $\xm$. If $g\geq
1$, the number of terms is infinite but the limit at $\xi\rightarrow
0^+$ is still finite.

The best way to find an expression for $\tKt$ is to use the residue
part of eq.~(\ref{3.2.3}). The sum over~(\ref{3.2.13}) can be
calculated with the help of eqs.~(\ref{3.1.13}) and~(\ref{3.1.32}):
\qq
\lefteqn{\hspace*{-.68in}
\tKt  =
\frac{(-1)^{n+g}\pi}{2^{2g-2}}
\left\{
\Res_{\phi=0}
\frac{\exp(2\pi i\phi\gamma)}{\denp}
\cot(\pi K\phi)
\right.
\pjn \sin\left[{2\pi\phi\over p_j}
\left(K\xnp_j + {\mu_j\over 2}\right)\right]
}
\nonumber\\
&&
-\fri\smpj\pmup
\left[\sum_{0\leq \xm <a} \frac{\sign{a}}{\spm{\xm}}
\right.
\label{3.2.15}\\
&&\qquad\left.\left.\left.
\times
\Res_{\phi=0}
\frac
{e^{2\pi i\gamma\phi} \exp[2\pi i\phi\sign{a}(m-|a|)]}
{\denp}\right]\right|_{a=-{\gamma\over K} + \sjnmuk}
\right\}
\nonumber
\qqq
It is clear from this formula that $\tKt$ is indeed a local
polynomial in $\gamma$, the breaks at the points~(\ref{3.2.12}) come
from the ``singular'' sum $\sum_{0\leq m<a}$.

The calculation of the integral over $\gamma$ in eq.~(\ref{3.2.10})
is now straightforward (but tedious). The integral is a sum of the
contributions of the stationary phase point $\gamma=0$ associated
with irreducible connections and break points~(\ref{3.2.12})
associated with reducible connections. To calculate the former one
has to take the polynomial which is equal to $\tKt$ in the vicinity
of $\gamma=0$ and substitute it in eq.~(\ref{3.2.10}) instead of
$\tKt$. To calculate a contribution of a point~(\ref{3.2.12}) one may
substitute the term of the sum $\sum_{\xm\in\ZZ}$ of
eq.~(\ref{3.2.11}) which has the break at that point, in the similar
way. These calculations lead ultimately to
eqs.~(\ref{3.1.37})-(\ref{3.1.40}). We do not discuss them here but
the examples for the case of a 3-fibered rational homology sphere
$X_0\left({p_1\over q_1},{p_2\over q_2},{p_3\over q_3}\right)$ can be
found in~\cite{Ro1}.

As we see, the residue calculations of Section~\ref{s3} are simpler
and more straightforward. However the calculations involving the
Kostant partition function present a clear group theoretical picture
by relating the surgery formula to multiplicities of irreducible
representations in tensor products of representations of quantum
groups (for more details see~\cite{Ro1}). This simplifies the
analysis of reducibility of connections providing the
contributions to Witten's invariant based on general simple Lie
groups.

\nappendix{2}
\label{a2}

In Section~\ref{s4} we used the fact that the moduli space $\mts$ of
flat connections on a punctured surface is a bundle over the moduli
space $\ms$. However the space $\ms$ is singular. Its singularity
results in the ``ugly'' sums like
$\sum_{0\leq\xxm<\left|\smut\right|}$ in eq.~(\ref{5.1.4}) and in the
requirement that the sums $\sjn\pm\theta_j$ should not be integer.

In order to avoid the singularity of $\ms$ E.~Witten suggested
in~\cite{Wi3} to consider the twisted $SO(3)$ bundle over $\Sigma_g$
for which the moduli space of flat connections is nonsingular. Since
we are dealing with punctured surfaces, we may even avoid using
$SO(3)$ directly although our formulas will be very similar to those
of~\cite{Wi3}.

The base for our bundles is the moduli space $\tms$ of flat
connections on $\Sigma_g$ with one puncture, the holonomy around
which is equal to $e^{i\pi\sigma_3}$. Note that $\dim\tms = \dim\ms
=3g-3$. In fact, $\tms$ is a $2^g$ times folded covering of the
moduli space of flat connections on the twisted $SO(3)$ bundle over
$\Sigma_g$. For the set of phases
\qq
\theta_1 = {1\over 2} - \ttheta_1,\;
\theta_2 = \ttheta_2,\;\ldots,\;
\theta_n = \ttheta_n;\qquad \ttheta_j\ll 1,
\label{5.2.1}
\qqq
the moduli space $\mts$ which we will also denote as  simply as
$\tmns$, is a bundle over $\tms$ in much the same way as it was a
bundle over $\ms$ when $\theta_1$ rather than $\ttheta_1$ was very
small. The reason why we can use notation $\tmns$ for $\mts$ is that
in contrast to the case of $\theta_j\ll 1$, the topological
class of $\mts$ does not depend on the phases $\theta_j$ as long as
$\ttheta_j\ll 1$.

Rewriting the r.h.s. of eq.~(\ref{2.1.12}) in terms of $\ttheta_j$ we
find that
\qq
\zt = - {1\over 2^{g-1}\pi^{n+2g-2}}
\sum_{\b\geq 1} (-1)^{\b}
\frac{e^{-a\b^2}}{\b^{n+2g-2}}
\pjn\sin(2\pi\b\ttheta_j).
\label{5.2.2}
\qqq
The extra factor $(-1)^\b$ translates into shifting the summation
from integer to half-integer $\xxm$ in the Poisson resummation
formula:
\qq
\sum_{\b\in\ZZ} (-1)^\b \delta(\b-x) =
\sum_{\xxm\in\ZZ + {1\over 2}} e^{2\pi i\xxm x}.
\label{5.2.3}
\qqq
As a result, instead of eq.~(\ref{3.1.32}) we should use
\qq
\sum_{\xxm >0\atop \xxm\in \ZZ + {1\over 2}}
e^{2\pi i\b\xxm} =
{i\over 2} {1\over \sin(\pi\b)}.
\label{5.2.4}
\qqq
We can also drop the second sum in eq.~(\ref{3.1.31}) if $|a|<{1\over
2}$, which is indeed the case if $\ttheta_j\ll 1$. Thus we get
\qq
\ztirr =
{1\over 2^g \pi^{n+2g-3}}
\Res_{\phi=0}
\frac{e^{-a\phi^2}}{\phi^{n+2g-2}}
\frac{\pjn\sin(2\pi\phi\ttheta_j)}
{\sin(\pi\phi)}
\label{5.2.5}
\qqq
instead of eq.~(\ref{5.1.4}).

If we introduce a set of integer numbers $\ta_j$ related to $\a_j$
\qq
\a_1 = K - \ta_1,\;\a_2 = \ta_2,\;,\ldots,\;
\a_n = \ta_n,
\label{5.2.6}
\qqq
then apparently
\qq
\nga = - \left({K\over 2}\right)^{g-1}
\sum_{\b=1}^{K-1} (-1)^{\b}
\frac
{\pjn \sin\left( {\pi\over K} \b\ta_j\right)}
{\den}.
\label{5.2.7}
\qqq
As a result, if $\ta_j\ll K$, then instead of eq.~(\ref{5.1.2})
\qq
\nga = 4\pi \left({K\over 2}\right)^g
\Res_{\phi=0}
\frac
{\pjn \sin(2\pi\phi\ta_j)}
{\denphi}{1\over\sin(2\pi K\phi)}.
\label{5.2.8}
\qqq

Finally, if we introduce the new numbers
\qq
\xnp_1 = {p_1\over 2} - \txnp_1,\;
\xnp_2 = \txnp_2,\;\ldots,\;
\xnp_n = \txnp_n,
\label{5.2.08}
\qqq
and assume that $\txnp_j\ll p_j$, then eq.~(\ref{5.1.19}) can be
rewritten as
\qq
\zirr & = &
-
\frac{(-1)^{nl+r_1} i^{Kr_1p_1}\pi}
{\pjn \Szpm{\xnp_j\over p_j}}
\left({K\over 2}\right)^{g}
\frac{\sign{P}}{\sqrt{|P|}} \eis
\label{5.2.9}\\
&&\times
\txep
\nonumber\\
&&\times
\ecass
\nonumber\\
&&\times
\smj \pmj
\tyep
%
%
\nonumber\\
&&\times
\Res_{\phi=0}
\frac{\zeph}
{\denp}
\frac
{\pjn \sin \left( 2\pi K\phi
\frac{\txnp_j + {\mu_j\over 2K}}
{p_j}
\right)}
{\sin(\pi K\phi)}
\nonumber
\qqq

Since eqs.~(\ref{5.1.18}) and~(\ref{5.1.16}) still hold:
\qq
&
2\ztirr = \int_{\tmns} \exp\left(\omega +
4a
\Theta
\right),\qquad\omega = \omega_0 + 2\sjn\ttheta_j\omega_j,
&
\label{5.2.10}\\
&
\nga = \int_{\tmns} \exp \left(K\omega_0 + \sjn\ta_j\omega_j\right)
\ahat(\tmns),
&
\label{5.2.11}
\qqq
we conjecture that
\qq
\lefteqn{
\hspace*{-1in}
\int_{\tmns} \exp \left[ K\left(\omega_0 + 2\sjn\ttheta_j\omega_j
\right) +
4a
\Theta \right]
\ahat(\tmns)
}
\label{5.2.12}\\
& = & 2\pi \left({K\over 2}\right)^g
\Res_{\phi=0}
\frac{e^{-a\phi^2}}{\denp}
\frac{\pjn\sin(2\pi K\phi\ttheta_j)} {\sin(\pi K\phi)}.
\nonumber
\qqq
We do not need to generalize this equation further to the analog of
eq.~(\ref{5.1.18}) because the manifold $\tmns$ is manifestly
independent of $\ttheta_j$.

Combining eqs.~(\ref{5.2.9}) and~(\ref{5.2.12}) we obtain the
nonsingular version of eq.~(\ref{5.1.20}) which holds for $\txnp_j\ll
p_j$:
\qq
\zirr & = &
-\frac{(-1)^{nl+r_1}i^{Kr_1 p_1}}
{2\pjn \Szpm{\xnp_j\over p_j}}
\frac{\sign{P}}{\sqrt{|P|}}\eis
\label{5.2.13}\\
&&\times
{\exp \left [ 2\pi iK \sjn
	\left( {r_j\over p_j} \tilde{\xn}_j^{\prime 2} -
	{1\over 4} s_j q_j l^2 \right)\right] }
\nonumber\\
&&\times
\ecass
\nonumber\\
&&\times
\smj \pmj
\tyep
\nonumber\\
&&\times
\int_{\tmns}
\exp \left[ K \left( \omega_0 + 2\sjn \frac{\txnp_j + {\mu_j\over 2K}}
{p_j} \omega_j +
2
\pi i{H\over P} \Theta \right)\right]
\ahat(\tmns).
\nonumber
\qqq


\begin{thebibliography}{99}
\bim{Wi1}{E. Witten}{Commun. Math. Phys.}{121}{1989}{351}
\bim{Je1}{L. Jeffrey}{Commun. Math. Phys.}{147}{1992}{563}
\bim{ReTu}{N. Reshetikhin, V. Turaev}{Invent. Math.}{103}{1991}{547}
\bim{FrGo}{D. Freed, R. Gompf}{Commun. Math. Phys.}{141}{1991}{79}
\bibitem{Ro1} L. Rozansky,
{\em A Large $k$ Asymptotics of Witten's Invariant of Seifert
Manifolds}, preprint UTTG-06-93, hep-th/9303099 (to be published in
Commun. Math. Phys.).
\bibitem{AxSi} S. Axelrod, I. Singer, {\em Chern-Simons Perturbation
Theory}, Proceedings of XXth Conference on Differential Geometric
Methods in Physics, Baruch College, C.U.N.Y., NY, NY.
\bibitem{BN1} D. Bar-Natan, {\em Perturbative Chern-Simons Theory},
Princeton Preprint, August 23, 1990.
\bim{Wi3}{E. Witten}{J. Geom. Phys.}{9}{1992}{303}
\bibitem{JeKi} L.~Jeffrey, F.~Kirwan, {\em Localization for
Nonabelian Group Actions}, Preprint, June 1993, (to be published in
Topology).
\bibitem{Sz2} A.~Szenes, {\em The Combinatorics of the Verlinde
Formulas}, preprint, alg-geom/9402003.
\bim{Wi2}{E.~Witten}{Comm. Math. Phys.}{141}{1991}{153}
\bim{Sz1}{A.~Szenes}{Topology}{32}{1993}{587}
\bim{BeSz}{A.~Bertram, A.~Szenes}{Topology}{32}{1993}{599}
\bibitem{Do}S.~K.~Donaldson, {\em Gluing Techniques in the Cohomology of
Moduli Spaces}, Topological Methods in Modern Mathematics, L.~Goldberg,
A.~Phillips (ed.), Publish or Perish Inc., Houston, 1993.
\bibitem{BlTh}M.~Blau, R.~Thompson, {\em Equivariant K\"{a}hler
Geometry and Localization in the $G/G$ Model}, preprint IC/94/108,
ENSLAPP-L-469/94, June 1994.
\bibitem{Ro2}L. Rozansky, {\em A Contribution of the Trivial
Connection to
the Jones Polynomial and Witten's Invariant of 3d Manifolds
I.}, preprint UMTG-172-93, UTTG-30-93, hep-th/9401061.
\end{thebibliography}
\end{document}